\documentclass[prb,twocolumn,showpacs,amsmath,amssymb]{revtex4}

\usepackage{graphicx}
\usepackage{dcolumn}
\usepackage{bm}
\usepackage{epsfig}

\begin{document}

\newcommand{\ep}{\varepsilon}
\newcommand{\up}{\uparrow}
\newcommand{\dn}{\downarrow}
\newcommand{\vectg}[1]{\mbox{\boldmath ${#1}$}}
\newcommand{\vect}[1]{{\bf #1}}

\title{Imaging mesoscopic spin Hall flow: Spatial distribution of local spin currents and spin densities in and out of multiterminal spin-orbit coupled semiconductor nanostructures}

\author{Branislav K. Nikoli\' c}
\author{Liviu P. Z\^ arbo}
\author{Satofumi Souma}
\altaffiliation[Present address: ]{Department of Physics, Faculty of Science, Tokyo University of Science,
1-3 Kagurazaka, Shinjuku-ku, Tokyo 162-8601, Japan}
\affiliation{Department of Physics and Astronomy, University
of Delaware, Newark, DE 19716-2570, USA}

\begin{abstract}
We introduce the concept of bond spin current, which describes the spin transport between two sites 
of the lattice model of a multiterminal spin-orbit (SO) coupled semiconductor nanostructure, and express it in 
terms of the spin-dependent nonequilibrium (Landauer-Keldysh) Green functions of the device. This formalism 
is applied to obtain the spatial distribution of microscopic spin currents in {\em clean}  phase-coherent two-dimensional electron gas with the Rashba-type of SO coupling attached to four external leads. Together with the corresponding profiles of the stationary spin density, such visualization of the phase-coherent spin flow allow us to resolve several key issues for the  understanding of mechanisms which generate pure spin Hall 
currents in the transverse leads of ballistic devices due to the flow of unpolarized charge current through their longitudinal leads: (i) while bond spin currents are non-zero locally within the SO coupled  region even in equilibrium (when all leads are  at the same potential), the total spin current obtained by summing the bond spin  currents over an arbitrary cross section is zero so that no spin can actually be transported by such equilibrium currents; (ii) when device is brought into nonequilibrium steady current state by external voltage difference applied between the longitudinal leads, only the wave functions (or Green functions) around the Fermi energy contribute to the total spin current through a given transverse cross section; (iii) the total spin Hall current  is not conserved within the SO coupled region; however, it becomes conserved and physically well-defined quantity in the ideal leads where it is, furthermore, equal to the spin current obtained within the multiprobe Landauer-B\" uttiker scattering formalism. The local spin current profiles crucially depend on  whether the sample is smaller or greater than the spin precession length, thereby demonstrating its essential role as the characteristic mesoscale for the spin Hall effect in ballistic multiterminal semiconductor nanostructures. Although static spin-independent disorder reduces the magnitude of the total spin current in the leads, the bond spin currents continue to flow through the whole diffusive 2DEG sample, without being localized as edge spin currents 
around any of its boundaries.
\end{abstract}

\maketitle

Recent experimental observation of the spin Hall effect~\cite{kato2004a,wunderlich2005a} 
opens new avenues for the understanding of fundamental role which spin-orbit (SO) couplings~\cite{rashba_review,winkler_book} can play in transport and equilibrium properties 
of semiconductor structures. While SO coupling effects are tiny relativistic corrections for particles 
moving through electric fields in vacuum, they can be enhanced in solids by several orders of magnitude 
due to the interplay of crystal symmetry and strong crystalline  potential.~\cite{rashba_review,winkler_book} Furthermore, harnessing of  spin currents induced by the spin Hall effect offers  new possibilities for 
the envisioned all-electrical manipulation of spin for semiconductor  spintronics 
applications~\cite{zutic2004a} where electrical fields can access individual spins  
on short length and time scales.

The principal macroscopic phenomenological manifestation of the spin Hall effect is unique: the transverse spin current, which is pure in the sense  of not being accompanied by any net charge transport in the transverse 
direction, emerges as a response to conventional unpolarized charge current in the longitudinal direction through a paramagnetic system in the absence of any external magnetic field. When such current hits the sample boundary, it will deposit nonequilibrium  spin accumulation~\cite{dyakonov1971a,hirsch1999a,brataas2005a,nikolic_accumulation,onoda2005a} at the lateral edges of the sample attached to two longitudinal electrodes,~\cite{nikolic_accumulation,onoda2005a} as detected optically in recent breakthrough spin Hall  experiments.~\cite{kato2004a,wunderlich2005a} 

However, there are several apparently disconnected mechanisms capable 
of inducing the spin Hall currents. Nonetheless, they share the necessity for 
some type of SO interaction which couples the spin and charge transport. For example, 
impurities with SO interaction will deflect spin-$\uparrow$ (spin-$\downarrow$) 
conduction electrons predominantly to the left (right) in the scattering process, 
thereby generating the {\em extrinsic} transverse spin Hall current. The theory  of the 
 extrinsic spin Hall effect has  been around for several  decades,~\cite{dyakonov1971a,hirsch1999a,brataas2005a} and it has recently been 
revisited~\cite{engel2005a} to argue its major role in one of the two recent seminal 
experimental observations.~\cite{kato2004a}  However, the extrinsic effect, which crucially  
relies on the presence of impurities with skew-scattering and does not involve any SO 
coupling induced modification of the quasiparticle energy spectrum, is a rather 
small effect (unless one invokes particular enhancement mechanisms involving intrinsic SO coupling 
in the bulk crystal which contributes a SO term to the impurity potential~\cite{engel2005a})  
whose precise magnitude has been  tantalizingly hard to estimate.~\cite{bernevig2004a} 

Thus, a strong impetus for the revival of interest in the realm of the spin Hall effect(s) 
has ascended from recent predictions for substantial pure spin current flowing through: (a) infinite 
homogeneous SO coupled semiconductors in the clean limit,~\cite{murakami2003a,sinova2004a}  
where the strong SO coupling induces the spin-splitting of the quasiparticle energies;  
or (b) the electrodes of multiprobe finite-size mesoscopic nanostructures~\cite{nikolic_mesoshe,souma_ringshe,sheng2005a} 
made of such materials. However, the theory of the intrinsic spin Hall effect~\cite{murakami_review} 
in the bulk of infinite semiconductors is formulated in terms of  the spin current density which is not conserved~\cite{rashba_review} in a medium with SO coupling and, moreover, can be non-zero even 
in thermodynamic equilibrium.~\cite{rashba2003a} Reexamination of such unusual features of the 
intrinsic spin Hall current  have lead to two major controversies: (i) its dependence solely 
on the whole SO coupled Fermi sea and the spin-split band structure (i.e., the anomalous 
velocity due to the Berry curvature of the Bloch states~\cite{murakami2003a,shen2004a}), 
without any connection to nonequilibrium distribution function that characterizes standard 
charge transport in the longitudinal direction,~\cite{zhang2005a} have prompted arguments that such currents 
do not really transport spin or induce spin accumulation that would be useful for spintronics applications;~\cite{zhang2005a,rashba2004a} (ii) for linear in momentum SO couplings, 
such as the Rashba or linear Dresselhaus coupling~\cite{winkler_book} in a two-dimensional 
electron gas (2DEG), numerous perturbative analytical~\cite{inoue2004a,malshukov2005a} and non-perturbative 
numerical exact diagonalization~\cite{sheng2005b} studies find that bulk intrinsic spin 
Hall current (averaged over an infinite system~\cite{shekhter2005a}) vanishes for arbitrary small 
disorder, while being able to survive as edge spin current near the sample-electrode interfaces~\cite{shytov2004a} or for some intermediate values of the spin-independent disorder strength~\cite{sugimoto2005a} (for which the sharp boundary between the intrinsic and the extrinsic mechanisms becomes blurred~\cite{schliemann2004a}).

On the other hand, the mesoscopic spin Hall current, predicted to flow out of ballistic phase-coherent 
samples of the SO coupled semiconductors~\cite{nikolic_mesoshe,souma_ringshe,sheng2005a,hankiewicz2004a,li2005a,wu2005a} 
through the attached transverse ideal (i.e., free of spin and charge interactions) leads, is conserved throughout  the leads, depends only on the wave functions (or Green functions) at the Fermi surface (at low temperatures $T \rightarrow 0$), and it is resilient to rather large static disorder within the 
diffusive metallic regime.~\cite{nikolic_mesoshe,sheng2005a} However, the theory of the {\em mesoscopic} spin 
Hall  effect is formulated in terms of the Landauer-B\" uttiker-type  transmission formalism~\cite{baranger1989a} for spin 
currents,~\cite{nikolic_mesoshe,souma_ringshe,pareek2004a} which connects asymptotic scattering 
states in the leads without requiring any information  about the quantum-mechanical probabilities 
for spin and charge propagation between the two points inside the sample. Technically, to obtain 
the spin Hall current flowing through the leads of a multiterminal device, one only needs the 
spin-dependent retarded real-space Green function connecting the sites residing in different leads and no 
information about its values between the points within the sample.~\cite{nikolic_mesoshe}

Thus, many recent debates on the very existence of the spin Hall effect in ballistic SO coupled 
semiconductor systems could be closed by visualizing the spatial details of the spin flow 
through experimentally accessible Hall bridges---from the SO coupled sample toward to attached 
electrodes with no SO interactions. The analogous studies of the spatial distribution   
of charge flow were essential in understanding the nature of  quantum Hall transport 
(bulk vs. edge~\cite{halperin1982a}) in mesoscopic Hall bridges.~\cite{gagel1995a,cresti2004a} Furthermore, 
recent advances  in multifarious scanning probe experimental techniques have made it possible 
to go beyond conventional transport measurements of macroscopically averaged quantities and 
image phase-coherent charge flow through a single 2DEG in quantum Hall or quantum 
point contact regime where the host semiconductor heterostructures is subjected 
to high or zero external magnetic field, respectively.~\cite{topinka2003a}

In particular, this type of insight can be obtained efficiently within the framework of  lattice 
models of mesoscopic devices and the corresponding bond charge currents~\cite{baranger1989a,todorov2002a} 
which yield a detailed picture of the charge propagation between two arbitrary sites of the lattice.~\cite{nonoyama1998a,cresti2003a} Here we provide a tool that makes it possible to obtain the spatial details of the spin flow on the scale of few nanometers by introducing the {\em bond spin currents}, which represent the analog of bond charge currents (and a lattice version of the spin current density). 

As shown in Sec.~\ref{sec:bond_spin}, the bond spin currents can be computed efficiently in terms of the spin-resolved  Keldysh Green functions~\cite{keldysh1965a} applied to the Landauer set-up  where a  finite-size SO coupled semiconductor sample is attached to many semi-infinite ideal leads.~\cite{caroli1971a,nonoyama1998a,cresti2003a} They are evaluated in Sec.~\ref{sec:ballistic}  for a paradigmatic mesoscopic spin Hall generator---a four-terminal ballistic two-dimensional electron  gas (2DEG) with the  Rashba type of SO coupling---to show the spatial profiles of spin currents and spin densities, thereby, revealing the features of the spin Hall transport on nanoscales. In Sec.~\ref{sec:disorder} we show that spatial distribution of the local spin current densities remains non-zero throughout the whole 2DEG even in the diffusive transport regime, in contrast to some previous conjectures~\cite{shytov2004a,shekhter2005a} where only the edge spin currents would survive the disorder effects in Rashba spin-split systems. The integration of the linear response bond spin currents over the transverse cross sections allows  us to connect in Sec.~\ref{sec:vs} the spin transport within the sample to the total spin Hall currents, which are obtained from the Landauer-B\" uttiker multiprobe spin current formulas, flowing  in an out  of the 2DEG through the leads as a response to the applied voltages at the device boundaries. We conclude in Sec.~\ref{sec:conclusion}.

\section{Bond Spin Currents in Multiterminal Nanostructures: Landauer-Keldysh Approach}\label{sec:bond_spin}

The conservation of charge implies the continuity equation in quantum mechanics for the charge density $\rho=e |\Psi({\bf r})|^2$  associated with a given wave function $\Psi({\bf r})$
\begin{equation} \label{eq:charge_continuity}
\frac{\partial \rho}{\partial t} + \nabla \cdot {\bf j} = 0, 
\end{equation}
from which one can extract the expression for the charge current density 
\begin{equation} \label{eq:charge_current}
{\bf j} = e {\rm Re} \, [ \Psi^{\dagger}({\bf r}) \hat{\bf v} \Psi({\bf r})].
\end{equation}
This can be viewed as the quantum-mechanical expectation value [in the state  $\Psi({\bf r})$] of the charge current density operator 
\begin{equation}\label{eq:charge_current_op}
\hat{\bf j} = e \frac{\hat{n}({\bf r})\hat{\bf v} + \hat{\bf v}\hat{n}({\bf r})}{2},
\end{equation}
which follows from the classical charge current density ${\bf j} = e n({\bf r}) {\bf v}$ via quantization procedure where the particle density $n({\bf r})$ and the velocity ${\bf v}$ are replaced by the corresponding operators and symmetrized to ensure that $\hat{\bf j}$ is a Hermitian operator.~\cite{baranger1989a} In SO coupled systems $\hat{\bf j}$ acquires extra terms since the velocity operator $i\hbar \hat{\bf v} = [\hat{\bf r},\hat{H}]$ is modified by the presence of SO terms in the Hamiltonian $\hat{H}$. For example, for the effective mass Rashba Hamiltonian of a finite-size 2DEG structure (in the $xy$-plane)
\begin{equation}\label{eq:rashba}
\hat{H} = \frac{\hat{\bf p}^2}{2m^*}  + \frac{\alpha}{\hbar} \left( \hat{p}_y \hat{\sigma}_x  - \hat{p}_x  \hat{\sigma}_y  \right) + V_{\rm conf}(x,y),
\end{equation}
the velocity operator is 
\begin{equation}\label{eq:velocity}
\hat{\bf v} = \frac{\hat{\bf p}}{m^*} - \frac{\alpha}{\hbar} (\hat{\sigma}_y {\bf e}_x - \hat{\sigma}_x {\bf e}_y),
\end{equation}
where  ${\bf e}_x$ and  ${\bf e}_y$ are the unit vectors along the $x$ and the $y$-axis, respectively.
Here $\hat{\bf p}=(\hat{p}_x,\hat{p}_y)$ is the momentum operator in 2D space, $\hat{\bm \sigma} =(\hat{\sigma}_x,\hat{\sigma}_y,\hat{\sigma}_z)$ is the vector of the Pauli  spin matrices, $\alpha$ 
is  the strength of the Rashba SO coupling~\cite{rashba_review,winkler_book} arising due to the 
structure inversion asymmetry,~\cite{pfeffer1998a} and $V_{\rm conf}(x,y)$ is the transverse confining potential. 

In contrast to the charge continuity equation Eq.~(\ref{eq:charge_continuity}), the analogous continuity equation for the spin density $\rho_s^i = \frac{\hbar}{2} [\Psi^\dagger ({\bf r}) \hat{\sigma}_i \Psi({\bf r})]$
\begin{equation}
\frac{\partial \rho_s^i}{\partial t} + \nabla \cdot {\bm {\mathcal J}}^i = S_s^i,
\end{equation}
contains the spin current density 
\begin{equation} \label{eq:spin_current_density}
{\bm {\mathcal J}}^i= \frac{\hbar}{2} \Psi^\dagger({\bf r}) \frac{\sigma_i \hat{\bf v} + \hat{\bf v} \sigma_i}{2}\Psi({\bf r}),
\end{equation}
as well as a non-zero spin source 
\begin{equation} \label{eq:spin_source}
S_s^i = \frac{\hbar}{2} {\rm Re} \, \left( \Psi^\dagger({\bf r}) \frac{i}{\hbar}[\hat{H},\hat{\sigma}_i] \Psi({\bf r}) \right). 
\end{equation}
The non-zero $S_s^i \neq 0$ term reflects non-conservation of spin in the presence of SO couplings. Thus, the plausible Hermitian operator of the spin current density~\cite{rashba_review} 
\begin{equation} \label{eq:spin_current_op}
\hat{\mathcal J}^i_k = \frac{\hbar}{2} \frac{\sigma_i \hat{v}_k + \hat{v}_k \sigma_i}{2}
\end{equation}
is a well-defined quantity (a tensor with nine components) only when the velocity operator is spin 
independent, as encountered in many metal spintronic devices.~\cite{brataas2001a} Such lack of 
physical justification for Eq.~(\ref{eq:spin_current_op}) in SO coupled systems leads to an arbitrariness~\cite{murakami2003a} in the definition of the spin current density employed 
in recent intrinsic spin Hall studies,~\cite{niu2005a} thereby casting a doubt on the experimental 
relevance of the quantitative predictions~\cite{murakami2003a,sinova2004a} for the spin Hall conductivity $\sigma_{sH}={\mathcal J}_y^z/E_x$ computed as the linear response to the applied longitudinal 
electric field $E_x$ penetrating an infinite SO coupled (perfect) semiconductor crystal.

To obtain the spatial profiles of spin and charge current densities in  finite-size samples of arbitrary shape attached to many probes, it is advantageous to represent the spin-dependent Hamiltonian and the 
corresponding charge and spin current density operators in the local orbital basis.~\cite{todorov2002a,caroli1971a,nonoyama1998a,cresti2003a} For example, in such representation the 
Rashba Hamiltonian can be recast in the following form~\cite{nikolic_accumulation}
\begin{eqnarray}\label{eq:tbh}
\hat{H}=\sum_{{\bf m}\sigma} \varepsilon_{\bf m} I_s \hat{c}_{{\bf
m}\sigma}^\dag\hat{c}_{{\bf m}\sigma}+\sum_{{\bf
mm'}\sigma\sigma'} \hat{c}_{{\bf m}\sigma}^\dag t_{\bf
mm'}^{\sigma\sigma'}\hat{c}_{{\bf m'}\sigma'},
\end{eqnarray}
where hard wall boundary conditions account for confinement on the lattice $L_x \times L_y$ with the lattice spacing $a$. Here  $\hat{c}_{{\bf m}\sigma}^\dag$ ($\hat{c}_{{\bf m}\sigma}$) is the creation (annihilation)  operator of an electron at the site ${\bf m}=(m_x,m_y)$.

While this Hamiltonian is of tight-binding type, its off-diagonal elements are non-trivial $2 \times 2$ Hermitian  matrices  ${\bf t}_{\bf m'm}=({\bf t}_{\bf mm'})^\dagger$ in the spin space. The on-site potential $\ep_{\bf m}$ describes any static local potential, such as the electrostatic potential due to the applied voltage or 
the disorder simulated via a uniform random variable $\varepsilon_{\bf m} \in [-W/2,W/2]$. The generalized 
nearest neighbor hopping $t^{\sigma\sigma'}_{\bf mm'}=({\bf t}_{\bf mm'})_{\sigma\sigma'}$ accounts for the 
Rashba coupling
\begin{eqnarray}\label{eq:hopping}
{\bf t}_{\bf mm'}=\left\{
\begin{array}{cc}
-t_{\rm o}{I}_s-it_{\rm SO}\hat{\sigma}_y &
({\bf m}={\bf m}'+{\bf e}_x)\\
-t_{\rm o}{I}_s+it_{\rm SO}\hat{\sigma}_x &  ({\bf m}={\bf m}'+{\bf e}_y)
\end{array}\right.,
\end{eqnarray}
through the SO hopping parameter $t_{\rm SO}=\alpha/2a$ ($I_s$ is the unit
$2 \times 2$ matrix in the spin space). The direct correspondence between the continuous Eq.~(\ref{eq:rashba}) and the lattice Hamiltonian Eq.~(\ref{eq:tbh}) is established by using  $t_{\rm o}=\hbar^2/2m^*a^2$ for the orbital hopping and by selecting the Fermi energy ($E_F=-3.8t_{\rm o}$ in the rest of the paper) at which zero-temperature quantum transport takes place close to the bottom of the band  $E_b=-4.0t_{\rm o}$ to ensure that injected  quasiparticles have quadratic and isotropic energy-momentum dispersion which characterizes the  Hamiltonians in effective mass approximation. Using the effective mass and conduction bandwidth of the semiconductor heterostructures measured in experiments,~\cite{nitta1997a} one can interpret the parameters of 
the lattice Hamiltonian as having the typical values $a \simeq 3$ nm for the lattice spacing while 
the Rashba SO hopping is of the order of  $t_{\rm SO}/t_{\rm o} \sim 0.01$.

The usage of the second quantized notation in Eq.~\ref{eq:tbh} facilitates the introduction of Keldysh Green function~\cite{keldysh1965a} expressions for the nonequilibrium expectation values.~\cite{caroli1971a,nonoyama1998a} We imagine that at time $t^\prime = -\infty$ the sample and the leads are not connected, while the left and the right longitudinal lead of a four-probe device are in their own thermal equilibrium with the chemical potentials $\mu_L$ and $\mu_R$, respectively, where $\mu_L=\mu_R+eV$. The adiabatic switching of the hopping parameter connecting the leads and the sample generates time evolution of the density matrix of the structure.~\cite{caroli1971a} The physical quantities are obtained as the nonequilibrium statistical average $\left< \ldots \right>$  (with respect to the density matrix~\cite{keldysh1965a} at time  $t^\prime=0$) of the corresponding quantum-mechanical operators expressed in terms of $\hat{c}_{{\bf m}\sigma}^\dag$ and $\hat{c}_{{\bf m}\sigma}$. This will lead to the expressions of the type 
$\left< \hat{c}^\dag_{{\bf m}\sigma} \hat{c}_{{\bf m}\sigma^\prime} \right>$, which define the lesser Green function~\cite{caroli1971a,nonoyama1998a} 
\begin{eqnarray} \label{eq:noneq_ev}
\left< \hat{c}^\dag_{{\bf m}\sigma}\hat{c}_{{\bf m}'\sigma^\prime} \right> &  = & \frac{\hbar}{i}G^<_{{\bf m}'{\bf m},\sigma^\prime \sigma}(\tau=0) \nonumber \\ 
&= &  \frac{1}{2\pi i}\int_{-\infty}^{\infty}dE G^<_{{\bf m}'{\bf m},\sigma^\prime \sigma}(E).
\end{eqnarray}
Here we utilize the fact that the two-time correlation  function [$\hat{c}_{{\bf m}\sigma}(t)=e^{i\hat{H}t/\hbar}\hat{c}_{{\bf
m}\sigma}e^{-i\hat{H}t/\hbar}$]
\begin{equation}\label{eq:lesser}
G^<_{{\bf m}{\bf m}',\sigma\sigma'}(t,t')\equiv\frac{i}{\hbar}\left<\hat{c}^\dag_{{\bf
m}'\sigma'}(t')\hat{c}_{{\bf m}\sigma}(t)\right>, 
\end{equation} 
depends only on $\tau=t-t^\prime$ in stationary situations, so it can be Fourier transformed to energy 
\begin{equation}
G^<_{{\bf m}{\bf m}',\sigma\sigma'}(\tau) = \frac{1}{2\pi\hbar}\int_{-\infty}^{\infty}dE G^<_{{\bf m}{\bf m}',\sigma\sigma'}(E)e^{iE\tau/\hbar},
\end{equation} 
which will be utilized for steady-state transport studied here. We use the notation where 
$\vect{G}^{<}_{\bf mm'}$ is a $2\times 2$ matrix in the spin space whose
$\sigma\sigma'$ element is $G^{<}_{{\bf mm'},\sigma\sigma'}$.

\subsection{Bond charge currents in SO coupled systems}

\subsubsection{Bond charge-current operator}
 
Using the charge conservation expressed through the familiar  
continuity equation Eq.~(\ref{eq:charge_continuity}) yields a 
uniquely determined bond charge current operator for quantum 
systems  described on a lattice by a tight-binding-type of Hamiltonian 
Eq.~(\ref{eq:tbh}). That is, the Heisenberg equation of motion 
\begin{eqnarray}\label{eq:heisenberg}
\frac{d\hat{N}_{\bf m}}{dt}=\frac{1}{i\hbar}\left[\hat{N}_{\bf
m},\hat{H}\right] .
\end{eqnarray}
for the electron number operator $\hat{N}_{\bf m}$ on site ${\bf m}$,
\begin{eqnarray}\label{eq:electrondensity}
\hat{N}_{\bf m}\equiv \sum_{\sigma=\up,\dn}\hat{c}_{{\bf
m}\sigma}^\dag\hat{c}_{{\bf m}\sigma},
\end{eqnarray}
leads to the charge continuity equation on the lattice 
\begin{eqnarray}\label{eq:continuity}
&&e\frac{d\hat{N}_{\bf m}}{dt}+\sum_{k=x,y}\left(\hat{J}_{{\bf m},{\bf m}+{\bf e}_k}-\hat{J}_{{\bf m}-{\bf e}_k,{\bf m}}\right)=0.
\end{eqnarray}
This equation introduces the bond charge-current operator~\cite{baranger1989a,todorov2002a}  $\hat{J}_{\bf mm'}$
which describes the particle current from site ${\bf m}$ to its nearest neighbor site ${\bf m}'$ (the `bond' terminology is supported by a picture where current between two sites is represented by a bundle of flow lines bunched together along a line joining  the two sites). 

Thus, the spin-dependent Hamiltonian Eq.~(\ref{eq:tbh}) containing the  non-trivial $2 \times 2$ matrix 
hopping defines the bond charge current operator $\hat{J}_{\bf mm'} =  \sum_{\sigma\sigma'} \hat{J}^{\sigma\sigma'}_{\bf mm'}$ which can be viewed as the sum of four different {\em spin-resolved} bond charge-current operators
\begin{equation}\label{eq:J_charge_op1_spinresolved}
\hat{J}^{\;\sigma\sigma'}_{\bf mm'} = 
\frac{e}{i\hbar}\left[\hat{c}_{{\bf m}'\sigma'}^\dag t_{ \bf
m'm}^{\sigma'\sigma}\hat{c}_{{\bf m}\sigma} -\mbox{h.c.}
\right],
\end{equation}
where h.c. stands for the Hermitian conjugate of the first term. In particular, for the case of 
$t_{ \bf mm'}^{\sigma\sigma'}$ being determined by the Rashba SO interaction Eq.~(\ref{eq:hopping}), we 
can decompose the bond charge current operator into two terms 
\begin{equation}\label{eq:J_charge_op2}
\hat{J}_{\bf mm'} = \hat{J}^{\rm kin}_{\bf
mm'}+\hat{J}^{\rm SO}_{\bf mm'},
\end{equation}
which have transparent physical interpretation. The first term
\begin{equation}\label{eq:J_charge_op_kinetic}
\hat{J}^{\rm kin}_{\bf mm'} =  \sum_{\sigma}
\frac{et_{\rm o}}{i\hbar} \left[ \hat{c}_{{\bf m}'\sigma}^\dag \hat{c}_{{\bf
m}\sigma} -\mbox{h.c.} \right]
\end{equation}
can be denoted as ``kinetic'' since it originates only from the kinetic energy 
$t_{\rm o}$ and does not depend on the SO coupling energy $t_{\rm SO}$, while the 
second term
\begin{eqnarray}\label{eq:J_charge_op_so}
\hat{J}^{\rm SO}_{\bf mm'}&=& \left\{
\begin{array}{cc}
\displaystyle{ -\frac{4et_{\rm SO}}{\hbar^2}\hat{S}_{\bf mm'}^{y}
} & ({\bf m}'={\bf m}+{\bf e}_x)\\
\displaystyle{ + \frac{4et_{\rm SO}}{\hbar^2}\hat{S}_{\bf mm'}^{x} }
& ({\bf m}'={\bf m}+{\bf e}_y)
\end{array}
\right.
\nonumber \\
&=& -\frac{4et_{\rm SO}}{\hbar^2}\left(({\bf m}'-{\bf
m})\times\hat{\vect{S}}_{\bf mm'}\right)_z
\end{eqnarray}
is the additional contribution to the intersite current flow due to 
non-zero Rashba SO hopping $t_{\rm SO}$. Here we also introduce the 
``bond spin-density'' operator 
\begin{eqnarray}\label{eq:bond_spindensity_op}
\hat{\vect{S}}_{\bf mm'}&=& \frac{\hbar}{4} \sum_{\alpha\beta}
\left[\hat{c}_{{\bf m'}\alpha}^\dag
\hat{\vectg{\sigma}}_{\alpha\beta}\hat{c}_{{\bf m}\beta} +\mbox{h.c.}
\right],
\end{eqnarray}
defined for the ``bond'' connecting the sites ${\bf m}$ and ${\bf m}'$, which 
reduces to the usual definition of the local spin density operator for 
${\bf m}'={\bf m}$ [see Eq.~(\ref{eq:spindensity})].

\subsubsection{Nonequilibrium bond charge current} \label{sec:neq_bc}

The formalism of bond charge currents yields the physically measurable~\cite{topinka2003a} 
local charge current within the sample as the quantum-statistical average  
$\left< \ldots \right>$ (with respect to a density matrix 
that has evolved over sufficiently long time so that nonequilibrium state and all interactions are 
fully established) of the bond charge-current operator in the nonequilibrium state,~\cite{caroli1971a,nonoyama1998a,cresti2003a}  
\begin{eqnarray}
\left<\hat{J}_{{\bf mm'}} \right>&=& \sum_{\sigma\sigma'}
\left<\hat{J}_{{\bf mm'}}^{\sigma \sigma'}\right>,\label{eq:J_charge_neq}
\\
 \left<\hat{J}_{{\bf
mm'}}^{\sigma\sigma'}\right>&=&\frac{-e}{\hbar}\int_{-\infty}^{\infty}
\frac{dE}{2\pi}\left[ t_{\bf m'm}^{\sigma'\sigma}G^{<}_{{\bf
mm'},\sigma\sigma'}(E) \right. \nonumber \\
&&\hspace{2cm}\left. -t_{\bf mm'}^{\sigma\sigma'}G^{<}_{{\bf
m'm},\sigma'\sigma}(E) \right],
\label{eq:J_charge_neq_spinresolved}
\end{eqnarray}
where we utilize Eq.~(\ref{eq:noneq_ev}) to express the local charge current in terms of the 
nonequilibrium lesser Green function. The spin-resolved bond charge current in (\ref{eq:J_charge_neq_spinresolved}) 
describes flow of charges which start as spin $\sigma$ electrons at the site ${\bf m}$ and end up as a 
spin $\sigma'$ electrons at the site ${\bf m}'$ where possible spin-flips  $\sigma \neq \sigma'$ (instantaneous or due to precession) are caused by spin-dependent interactions. The decomposition of the 
bond charge-current operator into kinetic and SO terms in  Eq.~(\ref{eq:J_charge_op2}) leads to a  
Green function expression for the corresponding nonequilibrium bond charge currents 
$\left<\hat{J}_{\bf mm'} \right> = \left<\hat{J}_{\bf mm'}^{\rm kin}\right>+\left<\hat{J}_{\bf mm'}^{\rm
SO}\right>$ with kinetic and SO terms given by
\begin{eqnarray}
\left<\hat{J}_{\bf mm'}^{\rm kin}\right>&=&
\frac{et_{\rm o}}{\hbar}\int_{-\infty}^{\infty} \frac{dE}{2\pi} {\rm
Tr}_s \left[ \vect{G}^{<}_{\bf mm'}(E) -\vect{G}^{<}_{\bf m'm}(E)
\right]\label{eq:J_charge_neq2_kin},
\nonumber \\ \\
\left<\hat{J}_{\bf mm'}^{\rm SO}\right>&=& \frac{et_{\rm
SO}}{\hbar}\int_{-\infty}^{\infty} \frac{dE}{2\pi i}\nonumber \\
&&\hspace{-1cm}\times {\rm Tr}_s \left[\left(({\bf m}'-{\bf
m})\times\vectg{\sigma}\right)_z\left( \vect{G}^{<}_{\bf
mm'}(E)+\vect{G}^{<}_{\bf m'm}(E) \right) \right]\label{eq:J_charge_neq2_so}. \nonumber \\
\end{eqnarray}
Note, however, that kinetic term is also influenced by the SO coupling through ${\bf G}^<$. 
In the absence of the SO coupling, Eq.~(\ref{eq:J_charge_neq2_so}) vanishes and the bond charge 
current reduces to the standard expression.~\cite{caroli1971a,nonoyama1998a,cresti2003a} The trace ${\rm Tr}_s$ 
here is performed in the spin Hilbert space. Similarly, we can also obtain the nonequilibrium 
local charge density in terms of ${\bf G}^<$
\begin{eqnarray} \label{eq:charge_density}
e\left<\hat{N}_{\bf m}\right> & = & e\sum_{\alpha=\up,\dn}
\left<\hat{c}^\dag_{{\bf m}\alpha}\hat{c}_{{\bf m}\alpha}\right> 
\nonumber \\ & = & \frac{e}{2\pi i}\int_{-\infty}^{\infty}dE
\sum_{\sigma}G^<_{{\bf m}{\bf m},\sigma\sigma}(E)
\nonumber \\
&=& \frac{e}{2\pi i}\int_{-\infty}^{\infty}dE \, {\rm Tr}_{\rm
s} [\vect{G}^<_{{\bf m}{\bf m}}(E)],
\end{eqnarray}
which is the statistical average value of the corresponding operator Eq.~(\ref{eq:electrondensity}).

\subsection{Bond Spin Currents in SO Coupled Systems}

\subsubsection{Bond spin-current operator}
To mimic the plausible definition of the spin-current density operator ${\mathcal J}^i_k$ 
in  Eq.~(\ref{eq:spin_current_op}), we can introduce the  bond
spin-current operator for the spin-$S_i$ component as the symmetrized product 
of the  spin-$\frac{1}{2}$  operator $\hbar \hat{\sigma}_i/2$  ($i=x,y,z$) and the bond 
charge-current operator from Eq.~({\ref{eq:continuity}}) 
\begin{eqnarray}\label{eq:bond_spincurrent_op1}
\hat{J}^{S_i}_{\bf mm'} &\equiv&\sum_{\alpha\beta}
\frac{1}{4i}\left[\hat{c}_{{\bf m}'\beta}^\dag
\left\{\hat{\sigma}_i,{\bf t}_{ \bf
m'm}\right\}_{\beta\alpha}\hat{c}_{{\bf m}\beta} -\mbox{h.c.}
\right].
\end{eqnarray}
By inserting the hopping matrix ${\bf t}_{ \bf m'm}$ Eq.~(\ref{eq:hopping}) of the lattice 
SO Hamiltonian into this  expression we obtain its explicit expression 
for the Rashba SO coupled system 
\begin{eqnarray}\label{eq:bond_spincurrent_op2}
\hat{J}^{S_i}_{\bf mm'}
 &=&\frac{t_{\rm o}}{2i}\sum_{\alpha\beta}\left(\hat{c}_{{\bf
m'}\beta}^\dag\left(\sigma_i\right)_{\beta\alpha} \hat{c}_{{\bf
m}\alpha}-\mbox{h.c.}\right) \nonumber
\\&&\hspace{1cm}+ \frac{t_{\rm
SO}}{2}\hat{N}_{\bf mm'}\left(\vect{e}_i\times
(\vect{m}'-\vect{m})\right)_z,
\end{eqnarray}
which can be considered as the lattice version of Eq.~(\ref{eq:spin_current_op}).
Here $\hat{N}_{\bf mm'}$ is the bond electron-number operator
\begin{eqnarray}
\hat{N}_{\bf mm'}\equiv \frac{1}{2}\sum_{\sigma}
\left(\hat{c}^\dag_{{\bf m}'\sigma}\hat{c}_{{\bf m}\sigma}+ {\rm
h.c.} \right),
\end{eqnarray}
which reduces to the standard electron-number operator
Eq.~(\ref{eq:electrondensity}) for ${\bf m}'={\bf m}$.

\subsubsection{Nonequilibrium bond spin current} \label{eq:neq_bs}

Similarly to the case of the nonequilibrium  bond charge current in Sec.~\ref{sec:neq_bc}, 
the  nonequilibrium statistical average of the bond spin-current operator Eq.~(\ref{eq:bond_spincurrent_op2}) 
can be expressed using the lesser Green function ${\bf G}^<$ as 
\begin{eqnarray}
\left<\hat{J}_{\bf mm'}^{S_i}\right>&=& \left<\hat{J}_{\bf
mm'}^{S_i{\rm (kin)}}\right>+ \left<\hat{J}_{\bf mm'}^{S_i{\rm
(so)}}\right>\label{eq:bond_spincurrent} \\
\left<\hat{J}_{\bf mm'}^{S_i{\rm (kin)}}\right>&=&
\frac{t_{\rm o}}{2}\int_{-\infty}^{\infty} \frac{dE}{2\pi} \nonumber
\\&&\hspace{0cm}\times{\rm Tr}\left[\sigma_i\left( \vect{G}^{<}_{\bf
mm'}(E)-\vect{G}^{<}_{\bf m'm}(E)\right)
\right],\label{eq:bond_spincurrent_kin}
\\
\left<\hat{J}_{\bf mm'}^{S_i{\rm (SO)}}\right>&=&
\left(\vect{e}_i\times (\vect{m}'-\vect{m})\right)_z \frac{t_{\rm
SO}}{2}\int_{-\infty}^{\infty} \frac{dE}{2\pi i} \nonumber
\\&&\hspace{0cm}\times {\rm Tr}\left[ \vect{G}^{<}_{\bf
mm'}(E)+\vect{G}^{<}_{\bf m'm}(E)
\right]\label{eq:bond_spincurrent_so}.
\end{eqnarray}
Here we also find two terms which can be interpreted as the kinetic and the SO 
contribution to  the bond spin current crossing from  site ${\bf m}$ 
to  site ${\bf m}'$. However, we emphasize that such SO contribution to 
the spin-$S_z$ bond current is identically equal to zero, which 
simplifies  the expression for this component to Eq.~(\ref{eq:bond_spincurrent_kin}) 
studied in the rest of the paper as the primary spin current response in the spin Hall effect. 

\subsubsection{Local Spin density and its continuity equation} \label{sec:local_spin}
The local spin density in the lattice models is determined by the 
local spin operator $\hat{{\bf S}}_{\bf m}=(\hat{S}^x_{\bf m},\hat{S}^y_{\bf
m},\hat{S}^z_{\bf m})$ at site ${\bf m}$ defined by
\begin{eqnarray}\label{eq:spindensity}
\hat{{\bf S}}_{\bf
m}=\frac{\hbar}{2}\sum_{\alpha\beta}\hat{c}^\dag_{{\bf
m}\alpha}\vectg{\sigma}_{\alpha\beta} \hat{c}_{{\bf m}\beta}.
\end{eqnarray}
The Heisenberg equation of motion for each component $\hat{\bf S}_i$ ($i=x,y,z$) 
of the spin density operator
\begin{equation}\label{eq:spinheisenberg}
\frac{d\hat{S}^i_{\bf m}}{dt}=\frac{1}{i\hbar}\left[\hat{S}^i_{\bf
m},\hat{H}\right]
\end{equation}
can be written in the following form
\begin{eqnarray}\label{eq:spincontinuity}
&&\frac{d\hat{S}^i_{\bf
m}}{dt}+\sum_{k=x,y}\left(\hat{J}^{S_i}_{{\bf m},{\bf m}+{\bf
e}_k}-\hat{J}^{S_i}_{{\bf m}-{\bf e}_k,{\bf
m}}\right)=\hat{F}_{\bf m}^{S_i},
\end{eqnarray}
where $\hat{J}^{S_i}_{{\bf m}{\bf m}'}$ is the bond spin-current
operator given by Eq.~(\ref{eq:bond_spincurrent_op1}) so that the
second term on the lefthand side of Eq.~(\ref{eq:spincontinuity})
corresponds to the ``divergence'' of the bond spin current on 
site ${\bf m}$. Here, in analogy with Eq.~(\ref{eq:spin_source}), 
we also find the lattice version of the spin source operator $\hat{F}_{\bf m}^{S_i}$ 
whose  explicit expression is given by
\begin{eqnarray}
\hat{F}_{\bf m}^{S_x}&=&-\frac{t_{\rm SO}}{t_{\rm o}}\left(\hat{J}^{S_z}_{{\bf
m},{\bf m}+{\bf e}_x}+\hat{J}^{S_z}_{{\bf m}-{\bf e}_x,{\bf m}}
\right),
\\
\hat{F}_{\bf m}^{S_y}&=&-\frac{t_{\rm SO}}{t_{\rm o}}\left(\hat{J}^{S_z}_{{\bf
m},{\bf m}+{\bf e}_y}+\hat{J}^{S_z}_{{\bf m}-{\bf e}_y,{\bf m}}
\right),
\\
\hat{F}_{\bf m}^{S_z}&=&\frac{t_{\rm SO}}{t_{\rm o}}\left(\hat{J}^{S_x}_{{\bf
m},{\bf m}+{\bf e}_x}+\hat{J}^{S_x}_{{\bf m}-{\bf e}_x,{\bf m}}
\right.\nonumber \\
&&\hspace{2cm}\left. +\hat{J}^{S_y}_{{\bf m},{\bf m}+{\bf
e}_y}+\hat{J}^{S_y}_{{\bf m}-{\bf e}_y,{\bf m}} \right).
\end{eqnarray}
The  non-zero term $\hat{F}_{\bf m}^{S_i}$ on the righthand side of  the spin continuity equation Eq.~(\ref{eq:spincontinuity}) is a formal expression, within the framework of bond spin current formalism, of 
the fact that spin is not conserved in SO coupled systems where it is forced into  precession by the 
effective momentum-dependent magnetic field of the  SO coupling. The fact that the bond spin  current operator Eq.~(\ref{eq:bond_spincurrent_op1}) appears in the spin continuity Eq.~(\ref{eq:spincontinuity}) as its divergence implies that its definition in Eq.~(\ref{eq:bond_spincurrent_op1}) is plausible. However, the presence of the spin source operator $\hat{F}_{\bf m}^{S_i}$ reminds us that such definition cannot be made unique,~\cite{niu2005a} unlike the case of the bond charge current which is uniquely  determined by the charge continuity Eq.~(\ref{eq:continuity}). 

If we evaluate the statistical average of Eq.~(\ref{eq:spincontinuity}) in a steady state (which can 
be either equilibrium or nonequilibrium), we obtain the identity
\begin{eqnarray}
\sum_{k=x,y}\left(\left<\hat{J}^{S_i}_{{\bf m},{\bf m}+{\bf
e}_k}\right>-\left<\hat{J}^{S_i}_{{\bf m}-{\bf e}_k,{\bf
m}}\right>\right)=\left<\hat{F}_{\bf m}^{S_i}\right>.
\end{eqnarray}
In particular, for the spin-$S_z$ component  we get
\begin{eqnarray}
&&\sum_{k=x,y}\left(\left<\hat{J}^{S_z}_{{\bf m},{\bf m}+{\bf
e}_k}\right>-\left<\hat{J}^{S_z}_{{\bf m}-{\bf e}_k,{\bf
m}}\right>\right)\nonumber \\
&&\hspace{0.5cm}=
\frac{t_{\rm SO}}{t_{\rm o}}\sum_{k=x,y}\left(\left<\hat{J}^{S_k}_{{\bf
m},{\bf m}+{\bf e}_k}\right>+\left<\hat{J}^{S_k}_{{\bf m}-{\bf
e}_k,{\bf m}}\right> \right),
\end{eqnarray}
which relates the divergence of the spin-$S_z$ current (lefthand side) to the spin-source (righthand side) determined by the sum of the longitudinal  component of the spin-$S_x$ current and the transverse component 
of the spin-$S_y$ current.

Since no experiment has been proposed to measure local spin current density within the SO 
coupled sample, defined through Eq.~(\ref{eq:spin_current_density}) or its lattice equivalent Eq.~(\ref{eq:bond_spincurrent}), we can obtain additional well-defined and measurable 
information about the  spin fluxes within the sample by computing the local spin density
\begin{eqnarray} \label{eq:spin_density}
\left<\hat{\vect{S}}_{\bf
m}\right> & = & \frac{\hbar}{2}\sum_{\alpha,\beta=\up,\dn}\vectg{\sigma}_{\alpha\beta}
\left<\hat{c}^\dag_{{\bf m}\alpha}\hat{c}_{{\bf m}\beta}\right>
\nonumber \\ & =  & \frac{\hbar}{4\pi
i}\int_{-\infty}^{\infty}dE\sum_{\alpha,\beta=\up,\dn}\vectg{\sigma}_{\alpha\beta}
G^<_{{\bf m}{\bf m},\alpha\beta}(E)
\nonumber \\
&=& \frac{\hbar}{4\pi i}\int_{-\infty}^{\infty}dE {\rm Tr}_{\rm
s}\left\{ \vectg{\sigma}\vect{G}^<_{{\bf m}{\bf m}}(E)\right\}.
\end{eqnarray}
Plotting of this quantity for Rashba SO coupled systems  will provides us with insight into the 
stationary spin flow in the nonequilibrium steady transport state.

\subsection{Spin-resolved Landauer-Keldysh Green functions for finite-size mesoscopic devices}
The formalism discussed thus far does not depend on the details 
of the external driving force which pushes the system into a 
nonequilibrium state. That is, the system can be driven by either the 
homogeneous electric field applied to an infinite homogeneous 2DEG or 
the voltage (i.e., electrochemical potential) difference between 
the electrodes attached to a finite-size {\em mesoscopic} sample. For example, 
in the latter case, the external bias voltage only shifts the relative chemical 
potentials of the reservoirs into which the longitudinal leads (employed to simplify 
the boundary conditions) eventually terminate, so that the electrons do  not feel any 
electric field in the course of ballistic propagation through clean 2DEG central region. The information 
about these different situations is encoded in the lesser Green  function ${\bf G}^<$.

Here we focus on experimentally accessible spin Hall bridges~\cite{nikolic_mesoshe} where finite-size 
central region (C),  defined on the $L \times L$ lattice, is attached to four external 
semi-infinite leads of the same width $L$. The leads at 
infinity terminate into the reservoirs where electrons are brought into thermal equilibrium, 
characterized by the Fermi-Dirac distribution function $f(E-eV_p)$, to  ensure the steady-state 
transport---in such Landauer setup~\cite{baranger1989a} current is limited by  quantum transmission 
through a potential profile while power is dissipated non-locally in the reservoirs. The voltage in 
each lead  of the four-terminal spin Hall bridge is $V_p$ ($p=1,\ldots,4$) so that the 
on-site potential $\varepsilon_{\bf m}$ within the leads has to  be shifted by $eV_p$.

The spin-dependent lesser Green function ${\bf G}^<$ defined in Eq.~(\ref{eq:lesser}) is evaluated within 
the finite-size sample region as a $2L^2 \times 2L^2$ matrix in the site$\otimes$spin space through the  spin-resolved  Keldysh equation for matrices~\cite{nikolic_accumulation}
\begin{eqnarray}\label{eq:keldysh}
{\bf G}^<(E)&=&{\bf G}(E){\bf \Sigma}^<(E){\bf G}^{\dag}(E),
\end{eqnarray}
which is valid in this form for steady-state transport when transients have died away.~\cite{keldysh1965a} 
Within the effective single-particle picture, the retarded Green function can be 
obtained by inverting the Hamiltonian
\begin{equation} \label{eq:retarded}
{\bf G}(E) =\left[E{\bf I}_C-{\bf H}_C-U_{\bf m}-\sum_{p}{\bf \Sigma}_{p}(E - eV_p)\right]^{-1},
\end{equation}
where the self-energies 
\begin{eqnarray} \label{eq:self_energies}
{\bf \Sigma}^<(E)&=&i \sum_{p}{\bf \Gamma}_{p}(E-eV_p)f(E-eV_p),
\\
{\bf \Gamma}_{p}(E)&=&i\left[{\bf \Sigma}_{p}(E)-{\bf
\Sigma}^\dag_{p}(E)\right],
\\
{\bf \Sigma}_{p}(E)&=&{\bf H}^{\dag}_{pC}\left[\left(E+i0_+\right){\bf I}_p-{\bf H}^{\rm
lead}_{p}\right]^{-1} {\bf H}_{pC},
\end{eqnarray}
are exactly computable in the non-interacting electron approximation and without any inelastic processes 
taking place within the sample. They take into account the ``interaction'' of the SO coupled sample with 
the attached leads in the Landauer transport setup where they generate finite that electrons spends within 
the 2DEG before escaping through the leads toward the macroscopic thermalizing reservoirs. Here ${\bf I}_C$ is the $2L^2\times 2L^2$ identity matrix, while ${\bf I}_{p}$ is the identity matrix in the infinite
site$\otimes$spin space of the $p$th lead, and we use the
following Hamiltonian matrices
\begin{eqnarray}
\left({\bf H}_{C}\right)_{{\bf m}{\bf
m}',\sigma\sigma'}&=&\left<1_{{\bf
m}\sigma}\right|\hat{H}\left|1_{{\bf m'}\sigma'}\right>,\;\;({\bf
m}, {\bf m'}\in C),
\nonumber \\
\left({\bf H}^{\rm lead}_{p}\right)_{{\bf m}{\bf
m}',\sigma\sigma'}&=&\left<1_{{\bf
m}\sigma}\right|\hat{H}\left|1_{{\bf m'}\sigma'}\right>,\;\;({\bf
m}, {\bf m'}\in p),
\nonumber \\
\left({\bf H}_{pC}\right)_{{\bf m}{\bf
m}',\sigma\sigma'}&=&\left<1_{{\bf
m}\sigma}\right|\hat{H}\left|1_{{\bf m'}\sigma'}\right>,\;\;({\bf
m}\in p, {\bf m'}\in\mbox{C}),\nonumber \\
\end{eqnarray}
with $\left|1_{{\bf m}\sigma}\right>$ being a vector in the Fock
space (meaning that the occupation number is one for the single
particle state $\left|{\bf m}\sigma\right>$ and zero otherwise) 
and $\hat{H}$ is the Hamiltonian given in
Eq.~(\ref{eq:tbh}).

In the general case of arbitrary applied bias voltage, the gauge invariance of 
measurable quantities (such as the current-voltage characteristic) with respect to the shift of electric potential everywhere by a constant $V$, $V_p \rightarrow V_p + V$ and $U_{\bf m} \rightarrow U_{\bf m} + V$,   is satisfied on the proviso that the retarded self-energies ${\bm \Sigma}_p(E-eV_p)$ introduced by each lead depend explicitly on  the applied voltages at the sample boundary, while the computation of the retarded Green function ${\bf G}(E)$  has to include the  electric potential landscape $U_{\bf m}$ within the sample~\cite{christensen1996a} [which can be obtained from the Poisson equation with charge density Eq.~(\ref{eq:charge_density})].  However, when the applied bias is low, so that linear response zero-temperature quantum transport takes place through  the sample [as determined by ${\bf G}(E_F)$], the exact profile  of the internal potential becomes irrelevant.~\cite{baranger1989a,nikolic1999a}

\section{Spatial distribution of local spin currents and spin densities in ballistic four-terminal Rashba  
SO coupled nanostructures} \label{sec:ballistic}

Under the time reversal transformation $t \rightarrow -t$, the mass, charge, and energy are even while the 
velocity and Pauli matrices are odd. Thus, the charge currents are $t$-odd and must vanish in thermodynamic equilibrium (in the absence of magnetic field). On the other hand, the spin current-density operator
Eq.~(\ref{eq:spin_current_op}), which is the product of two $t$-even quantities, can have its 
expectation values to be non-zero even in thermodynamic equilibrium. This has been 
explicitly demonstrated~\cite{rashba2003a} for its expectation values in the eigenstates  
of an infinite clean Rashba spin-split 2DEG.~\cite{rashba2003a} 

\begin{figure*}
\centerline{\psfig{file=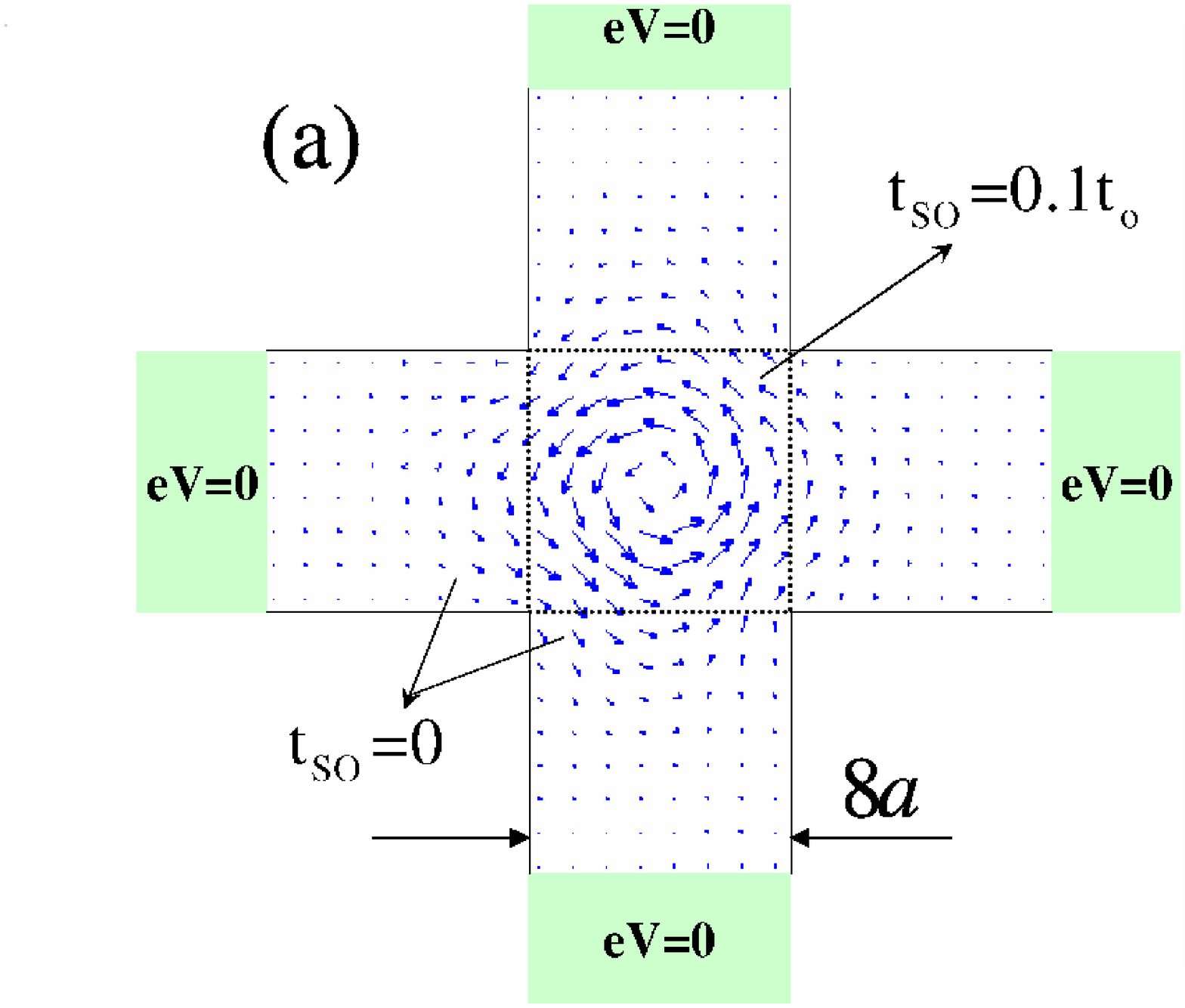,scale=0.32,angle=-90} \hspace{0.5in} \psfig{file=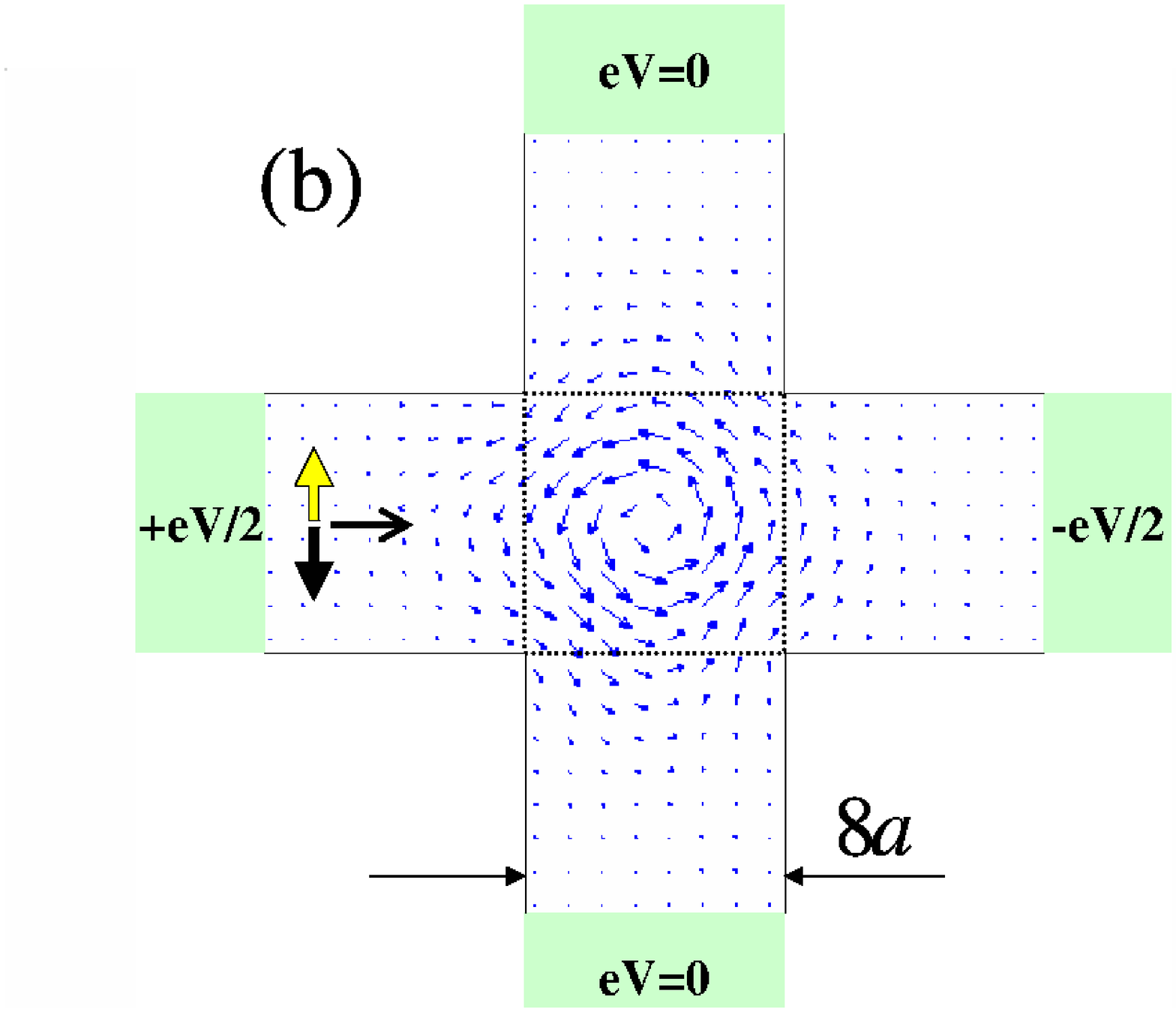,scale=0.32,angle=-90}} \vspace{0.5in}
 \centerline{\psfig{file=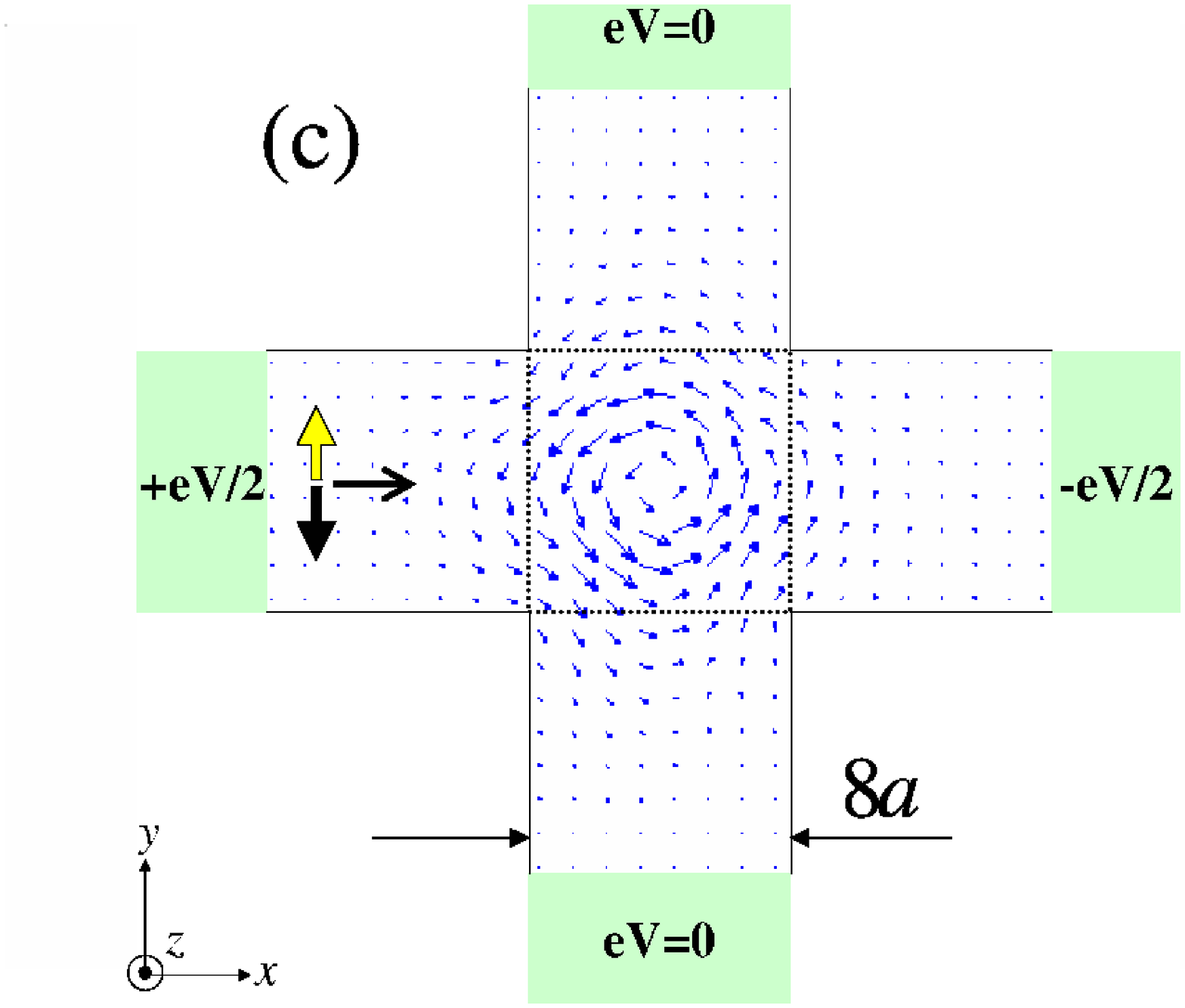,scale=0.32,angle=-90} \hspace{0.5in} \psfig{file=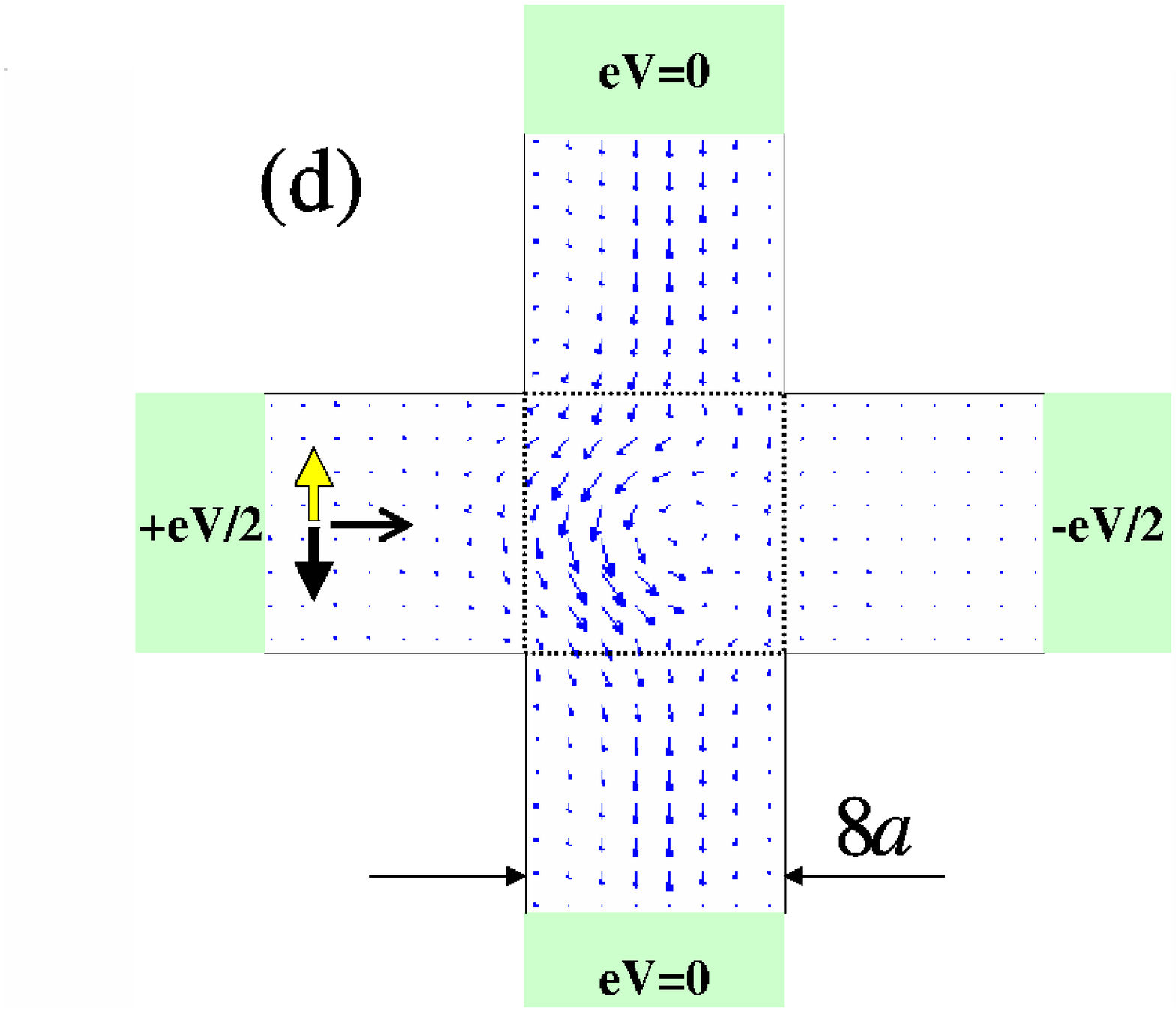,scale=0.32,angle=-90}}
\caption{(Color online) The spatial distribution of the local spin current in ballistic four-terminal bridges where the central 2DEG region, with the Rashba SO coupling $t_{\rm SO}=0.1t_{\rm o}$ and the corresponding spin precession length $L_{\rm SO} = \pi t_{\rm o}a/2t_{\rm SO} \approx 15.7a$ (typically $a \simeq 3$ nm), is attached to four ideal ($t_{\rm SO} \equiv 0$) leads. The magnitude of the bond spin current is proportional to the length of the arrow. The device is in equilibrium in (a), and out of equilibrium in (b), (c), and (d) due to the applied  bias voltage $eV =10^{-3} t_{\rm o}$ which drives the linear response longitudinal charge current and the transverse spin Hall current induced by its passage through the SO coupled region. The local spin current in (b), which is ``carried'' by all states from $-4t_{\rm o}$ (band bottom) to $E_F + eV/2$ ($E_F = -3.8t_{\rm o}$), is the sum of the equilibrium (persistent) spin current $\left< \hat{J}_{\bf mm'}^{S_z({\rm eq})} \right>$ in (c), carried by the fully occupied states from $-4t_{\rm o}$ to $E_F - eV/2$, and the nonequilibrium (transport)  spin current $\left< \hat{J}_{\bf mm'}^{S_z({\rm neq})} \right>$ in (d) carried by the partially occupied states around the Fermi energy from $\mu_R=E_F - eV/2$ (electrochemical potential of the right reservoir) to  $\mu_L=E_F + eV/2$ (electrochemical potential of the left reservoir).}\label{fig:eq_vs_neq}
\end{figure*}

To investigate possible patterns of such equilibrium local spin currents in mesoscopic finite-size devices we plot 
in Fig.~\ref{fig:eq_vs_neq}(a) the spatial distribution of the bond spin  currents, carried by the whole 
Fermi sea, in a four-terminal ballistic device with no impurities where all leads are kept at the same potential. Although we find  non-zero local spin currents, they {\em do not} transport any spin since the 
{\em total} spin current,  obtained by summing the bond spin currents over an arbitrary transverse cross section of the device 
\begin{equation} \label{eq:total_trans}
I^s_{\rm trans}(m_y) = \sum_{m_x} \left <\hat{J}_{(m_x,m_y)(m_x,m_y+1)}^{S_z}\right>, 
\end{equation}
or over any longitudinal cross section 
\begin{equation} \label{eq:total_long}
I^s_{\rm long}(m_x) = \sum_{m_y} \left <\hat{J}_{(m_x,m_y)(m_x+1,m_y)}^{S_z}\right>, 
\end{equation}
is identically equal to zero $I^s_{\rm trans}(m_y) = I^s_{\rm long}(m_x) \equiv 0$. 
This picture also provides a direct microscopic proof that no equilibrium total spin currents, 
conjectured in Ref.~\onlinecite{pareek2004a}, can actually appear in the leads of an unbiased $V_p = {\rm const.}$ mesoscopic device in thermodynamic equilibrium, as demonstrated recently~\cite{souma_ringshe,kiselev2004a} 
within the Landauer-B\" uittker approach which operates only with the total spin and charge 
currents in the leads.

In Fig..~\ref{fig:eq_vs_neq}(b) we apply low (i.e., linear response, see Sec.~\ref{sec:vs}) bias voltage $eV=10^{-3} t_{\rm o} \ll (E_F - E_b)=0.2 t_{\rm o}$ between the longitudinal leads and integrate 
expression Eq.~(\ref{eq:bond_spincurrent_kin}) from the bottom  of the band to the chemical potential 
$E_F + eV/2$ of the left reservoir. In contrast to the equilibrium spin current density from Fig.~\ref{fig:eq_vs_neq}(a), the vortex pattern is now distorted and non-zero total spin current 
$I^s_{\rm trans}(m_y) \neq 0$ in Eq.~(\ref{eq:total_trans}) emerges in the transverse direction, as 
expected in the phenomenology of the spin Hall effect. 

One of the highly unconventional features of the intrinsic spin Hall current is its dependence on the 
whole SO coupled Fermi sea,~\cite{murakami2003a,sinova2004a} even when infinite system is driven out of equilibrium by the applied external electric field thereby limiting the charge transport and the extrinsic 
spin Hall response to the Fermi level through the nonequilibrium part of the distribution function.~\cite{zhang2005a} However, such property appears to be  alien to the spirit of Fermi liquid 
theory where transport involves only quasiparticles with energies within $k_B T$ of the Fermi level.  

While the spin-currents crossing the transverse bonds in Fig.~\ref{fig:eq_vs_neq}(b) are apparently carried 
by the whole Fermi sea, we now  separate the integration in Eq.~(\ref{eq:bond_spincurrent_kin}) for $S_z$ bond spin current into two parts 
\begin{widetext}
\begin{eqnarray} \label{eq:integration}
\left<\hat{J}_{\bf mm'}^{S_z}\right> & =  &
\frac{t_{\rm o}}{2}\int\limits_{E_b}^{E_F-eV/2} \frac{dE}{2\pi}{\rm Tr}\left[\hat{\sigma}_z \left( \vect{G}^{<}_{\bf mm'}(E)-\vect{G}^{<}_{\bf m'm}(E) \right)\right] + \frac{t_{\rm o}}{2}\int\limits_{E_F - eV/2}^{E_F + eV/2} \frac{dE}{2\pi} {\rm Tr}\left[\hat{\sigma}_z \left( \vect{G}^{<}_{\bf
mm'}(E)-\vect{G}^{<}_{\bf m'm}(E)\right)\right] \nonumber \\
& = & \left< \hat{J}_{\bf mm'}^{S_z({\rm eq})} \right> + \left< \hat{J}_{\bf mm'}^{S_z({\rm neq})}\right>.
\end{eqnarray}
\end{widetext}
The states from the band bottom $E_b$ to $E_F -eV/2$ are fully occupied, while states in the energy 
interval from the electrochemical potential $E_F-eV/2$ ($e<0$) of the right reservoir to the electrochemical potential $E_F+eV/2$ of the left reservoir are partially occupied because of the competition between the 
left reservoir  which tries to fill them and the right reservoir which tries to deplete them. The profile 
of the first term $\left<\hat{J}_{\bf mm'}^{S_z(\rm eq)}\right>$ in Eq.~(\ref{eq:integration}) is shown 
in Fig.~\ref{fig:eq_vs_neq}(c), while the spatial profile of the second term, representing the local spin current $\left<\hat{J}_{\bf mm'}^{S_z(\rm neq)}\right>$ carried by the states around the Fermi energy is 
shown in Fig.~\ref{fig:eq_vs_neq}(d). 

The spatial distribution of the microscopic spin currents in Fig.~\ref{fig:eq_vs_neq}(c) is akin to the 
vortex-like pattern of bond spin currents within the device in equilibrium in Fig.~\ref{fig:eq_vs_neq}(a), 
and, therefore, {\em does not} transport any spin between two points in real space. Thus, Fig.~\ref{fig:eq_vs_neq} convincingly 
demonstrates that  the  non-zero spin Hall  flux through the  transverse cross sections in Fig.~\ref{fig:eq_vs_neq}(b) and Fig.~\ref{fig:eq_vs_neq}(d) is due to only the wave functions 
(or Green functions) at the Fermi energy (as $T \rightarrow 0$), in accord with the general paradigms  
of the Landau's Fermi liquid  theory where transport quantities are expected to be expressed as the 
Fermi-surface property.~\cite{baranger1989a,haldane2004a} 

We recall here that similar situation appears in charge transport in an external magnetic field  where  equilibrium (or persistent) current density,~\cite{baranger1989a} or bond charge currents in the lattice formalism,~\cite{cresti2004a,cresti2003a} is non-zero even in  unbiased devices (all leads at the same potential) in thermal equilibrium due to breaking of time-reversal invariance by the external magnetic field. However, such circulating or diamagnetic currents carried  by the Fermi sea, which in Landauer-Keldysh formalism can be subtracted by separating the integration~\cite{cresti2004a} in a fashion similar to our  Eq.~(\ref{eq:integration}), do not contribute to the net charge transport (i.e., to the total charge current  measured in experiments) through any cross section  of the device.~\cite{baranger1989a,cresti2004a} Thus, early ``Fermi sea'' expressions for  the linear response  transport coefficients in, e.g., the quantum Hall effect theory~\cite{baranger1989a} 
or in the anomalous  Hall effect theory,~\cite{haldane2004a} were eventually recast in terms of the Fermi surface determined quantities. Similarly, Fig.~\ref{fig:eq_vs_neq} demonstrates that spin Hall current carried by the ``bulk'' of the  Fermi sea, which is equilibrium and does not really transport spin between two points in space, must  be subtracted~\cite{rashba2003a} within any sensible theory for the intrinsic spin Hall conductivity defined in the thermodynamic limit of macroscopic systems.

\begin{figure*}
\centerline{\psfig{file=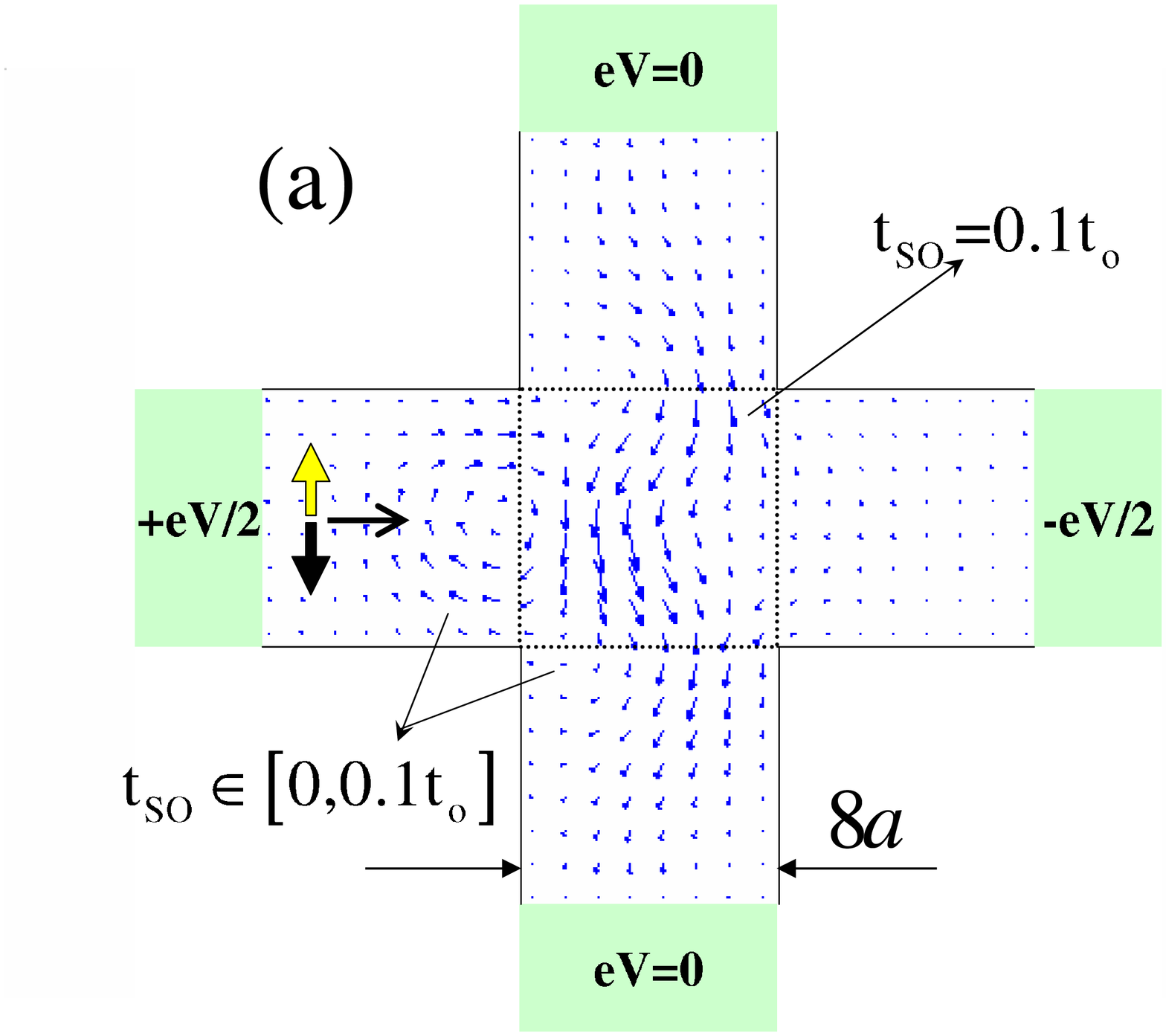,scale=0.32,angle=-90} \hspace{0.6in} \psfig{file=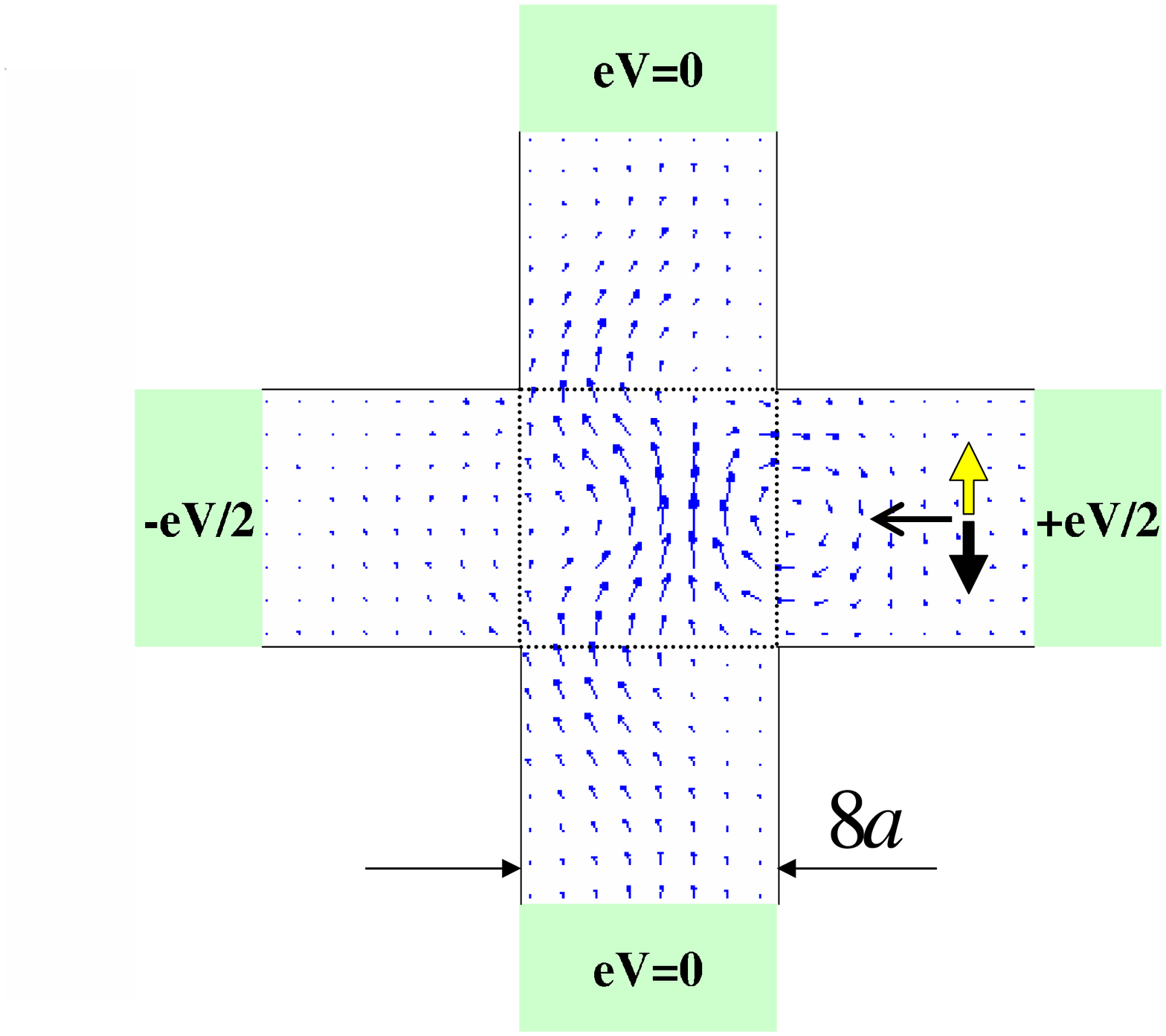,scale=0.32,angle=-90}} \vspace{0.4in}
 \centerline{\psfig{file=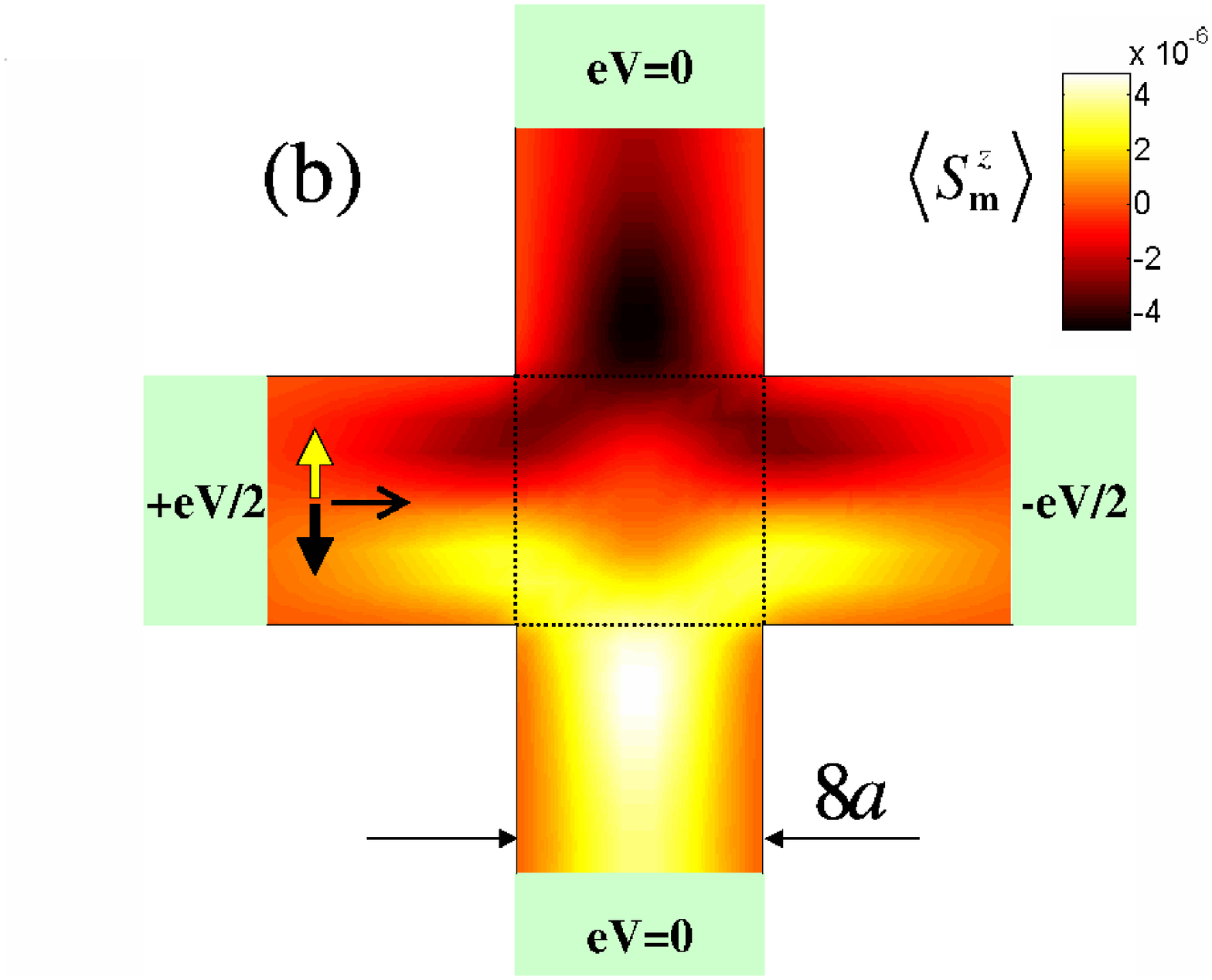,scale=0.32,angle=-90} \hspace{0.6in} \psfig{file=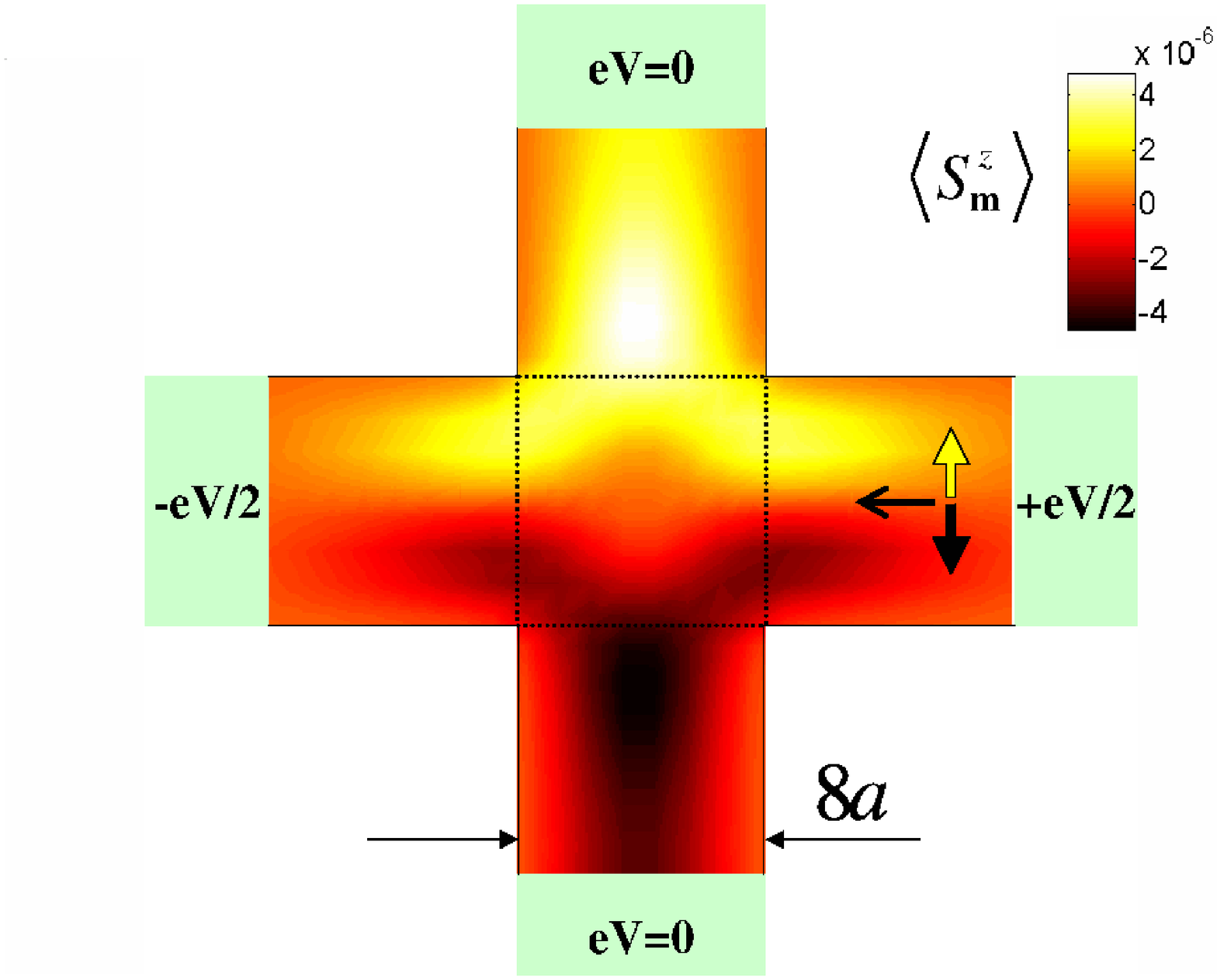,scale=0.32,angle=-90}} \vspace{0.4in}
 \centerline{\psfig{file=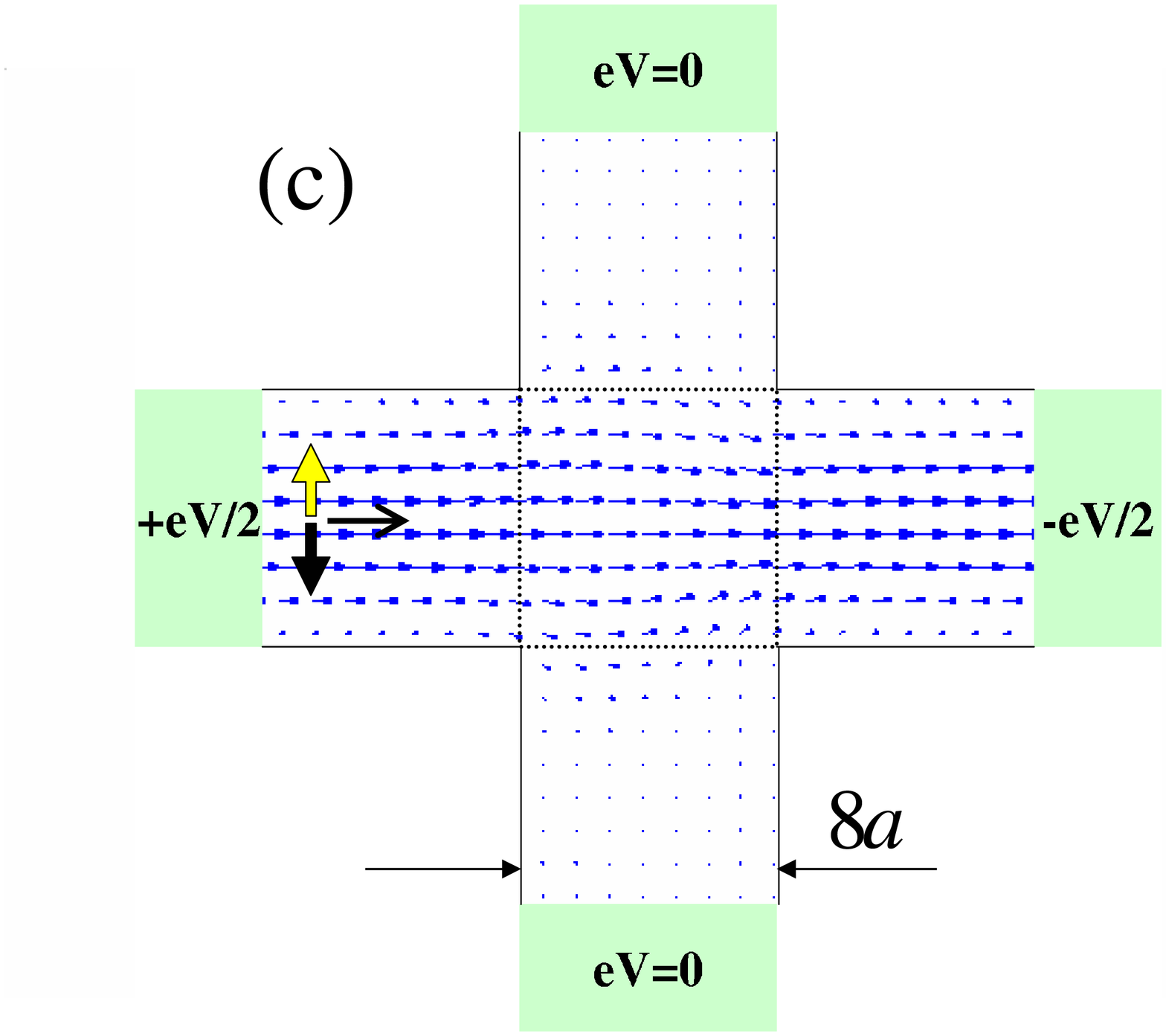,scale=0.32,angle=-90} \hspace{0.6in} \psfig{file=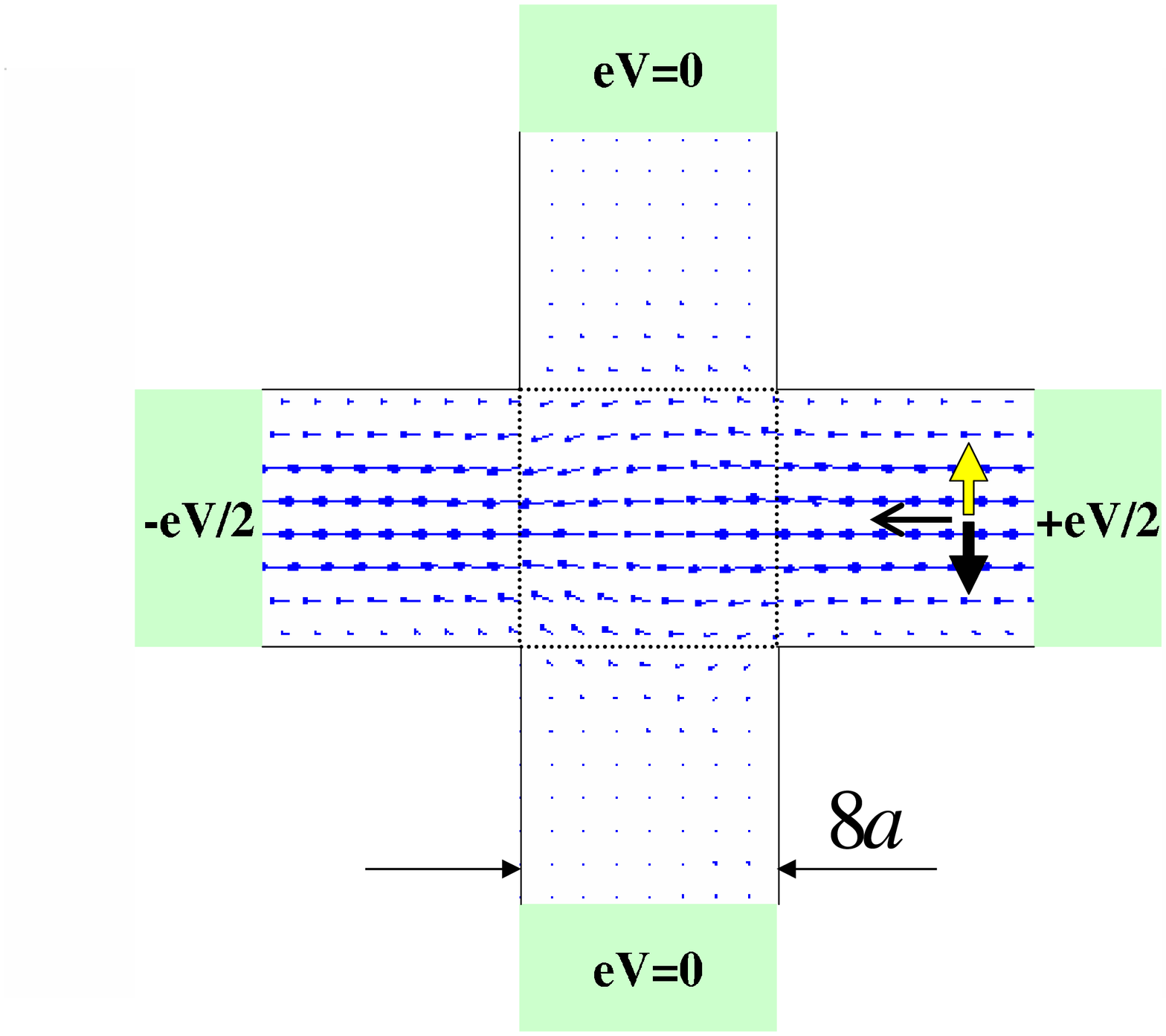,scale=0.32,angle=-90}}
\caption{(Color online) (a) The spatial distribution of the nonequilibrium local spin current $\left< \hat{J}_{\bf mm'}^{S_z({\rm neq})} \right>$ in ballistic four-terminal spin Hall bridges where the central finite-size 2DEG  (of size $8a \times 8a$) with the Rashba SO coupling $t_{\rm SO}=0.1t_{\rm o}$ ($L_{\rm SO} \approx 15.7a$) is attached to four leads containing a region (adjacent to the sample) of length $8a$ within which the SO coupling is switched on adiabatically (using a linear function) from $t_{\rm SO}=0$ to $t_{\rm SO}=0.1t_{\rm o}$. Panel (b) shows  the spatial profiles of the stationary flow of the local spin density $\left<\hat{S}_{\bf m}^z \right>$, while  panel (c) plots the spatial distribution of the local charge current density $\left< \hat{J}_{\bf mm'}  \right>$. Upon reversing the bias voltage $V \rightarrow -V$ driving the linear response ($eV=10^{-3}t_{\rm o}$), spatial profiles in the left column turn into the profiles of the right column where both the total longitudinal charge current and the total transverse spin Hall current change their direction.~\cite{wunderlich2005a}}\label{fig:adiabatic}
\end{figure*}

One of the basic tests for theories of the spin Hall effect is to predict the direction 
(i.e., the sign~\cite{engel2005a,nikolic_accumulation,niu2005a}) of the spin Hall 
current or the corresponding nonequilibrium spin Hall accumulation deposited by such 
current on the lateral boundaries of experimental devices.~\cite{kato2004a,wunderlich2005a}
 The nonequilibrium (linear response) spin Hall current in Fig.~\ref{fig:eq_vs_neq}(d) flows from 
 the top to the bottom transverse lead because the spin-$\uparrow$ electrons are deflected to the 
 right. This feature can be understood using the semiclassical picture based on the SO force operator~\cite{nikolic_soforce} generated by the Rashba Hamiltonian Eq.~(\ref{eq:rashba}) 
 of the finite-size 2DEG
\begin{equation}\label{eq:so_force}
\hat{\bf F}_{\rm SO} = \frac{2 \alpha^2 m^* }{\hbar^3}  (\hat{\bf p} \times {\bf
z}) \otimes \hat{\sigma}^z  - \frac{d V_{\rm conf}(\hat{y})}{d\hat{y}} {\bf y},
\end{equation}
and its expectation value in the spin-polarized wave packet state $|\Psi \rangle \otimes 
|\!\! \uparrow \rangle$. This apparently simple picture also explains why the transverse 
spin Hall current density bends toward the right in Fig.~\ref{fig:eq_vs_neq} while passing 
through the SO coupled region. 

However, this expectation value (i.e., the SO ``force'') oscillates along the sample due to the 
precession of the deflected spin in  the effective Rashba magnetic field which is nearly parallel to the $y$-axis because of the transverse confinement effects.~\cite{nikolic_soforce} In ballistic strongly 
coupled SO structures such $\alpha^2$-dependent SO ``force'', which  oscillates on the mesoscale set by the 
spin  precession length $L_{\rm SO} = \pi \hbar^2/ 2m^*\alpha = \pi t_{\rm o}a/2t_{\rm SO}$ (on which 
the spin precesses by an angle $\pi$, i.e., the state  $|\!\! \uparrow \rangle$ evolves into 
$|\!\! \downarrow \rangle$), will lead to a change of the sign~\cite{nikolic_mesoshe} of spin Hall current 
as a function of the system size $L/L_{\rm SO}$. Also, since the mesoscopic spin Hall effect sensitively depends on the measurement geometry,~\cite{nikolic_mesoshe} the sign of the transverse spin Hall current 
can change when non-ideal leads~\cite{sheng2005a} are attached to the sample.

\begin{figure*}
\centerline{\psfig{file=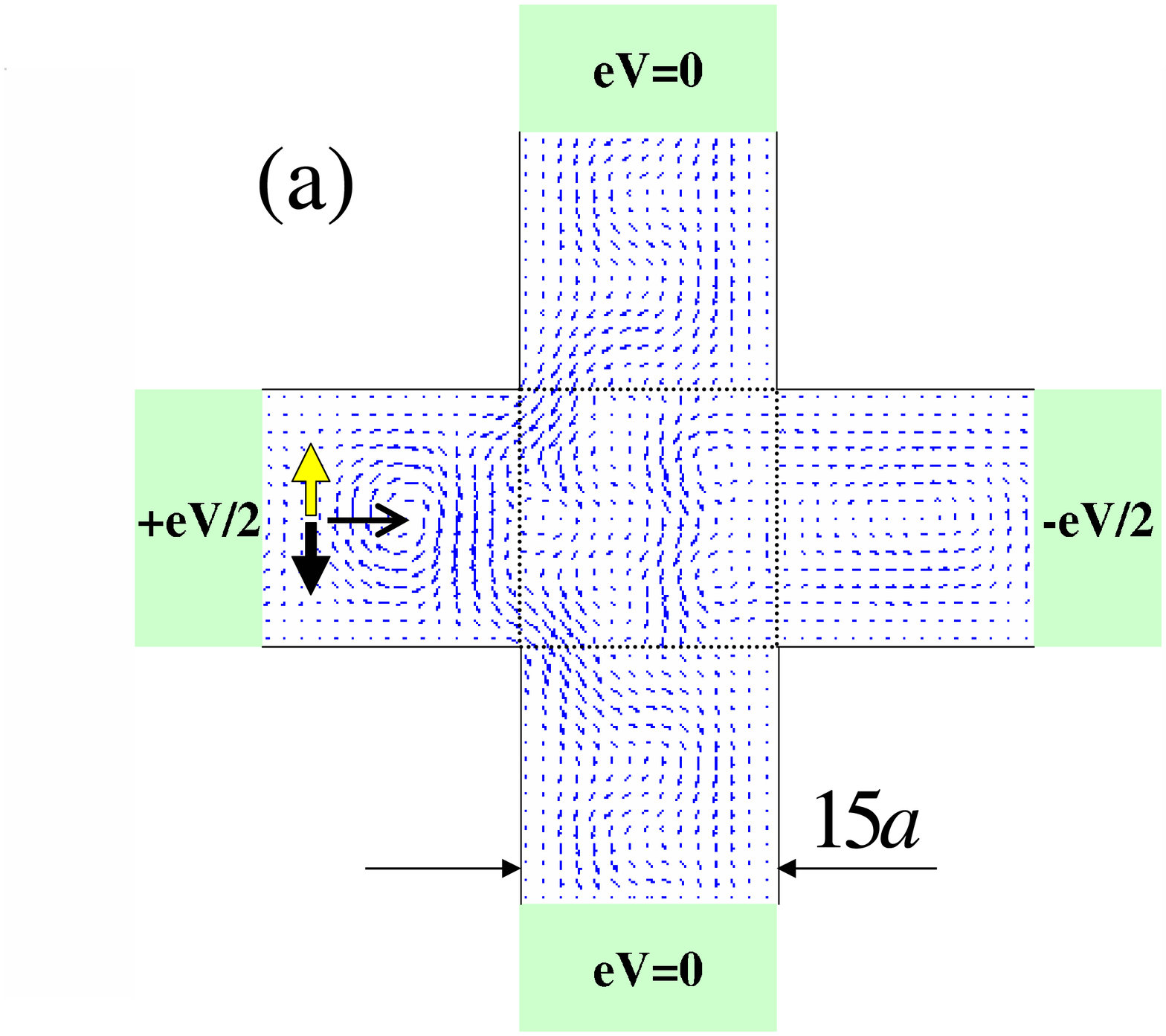,scale=0.38,angle=-90} \hspace{0.5in} \psfig{file=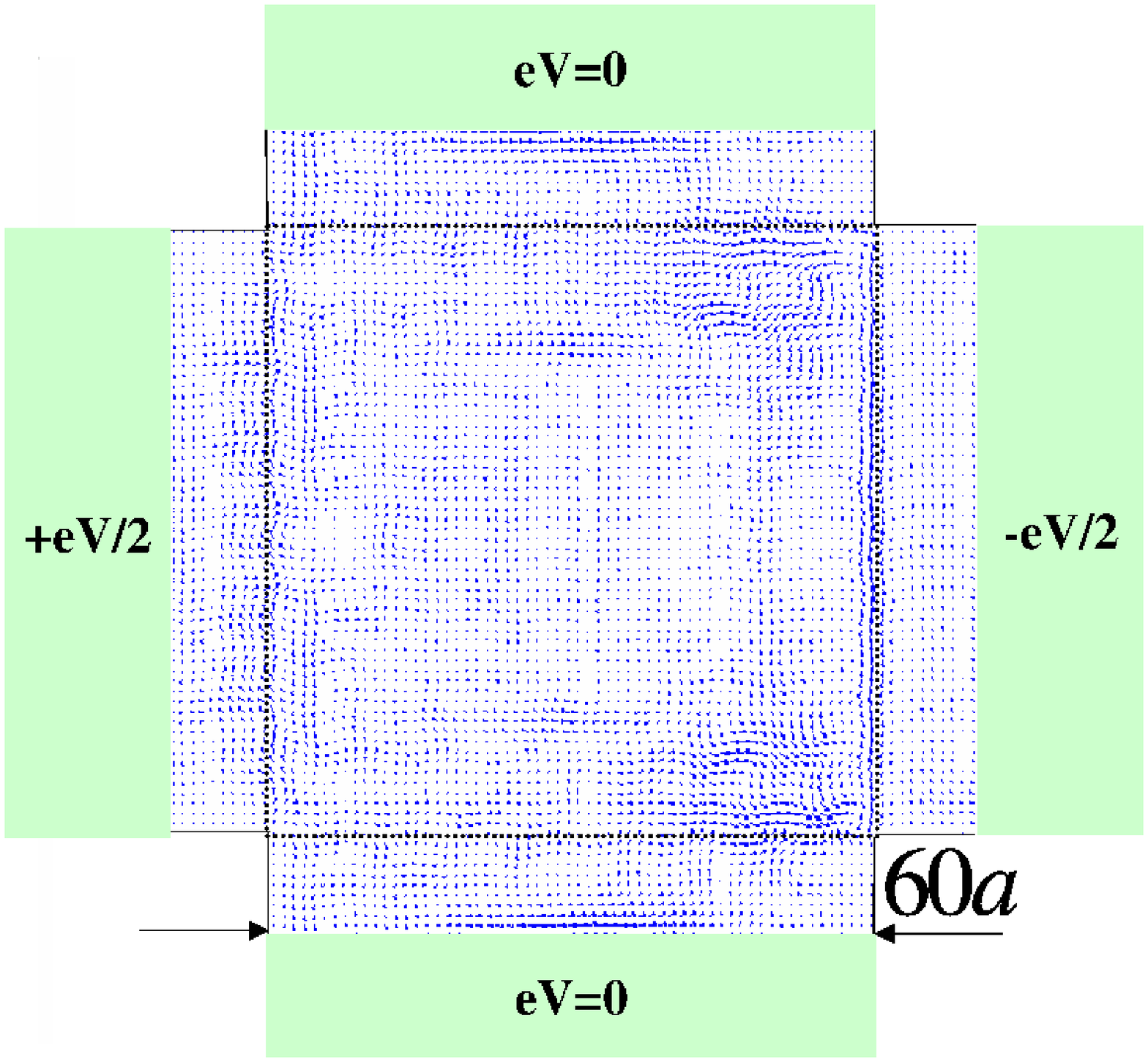,scale=0.38,angle=-90}} \vspace{0.5in}
 \centerline{\psfig{file=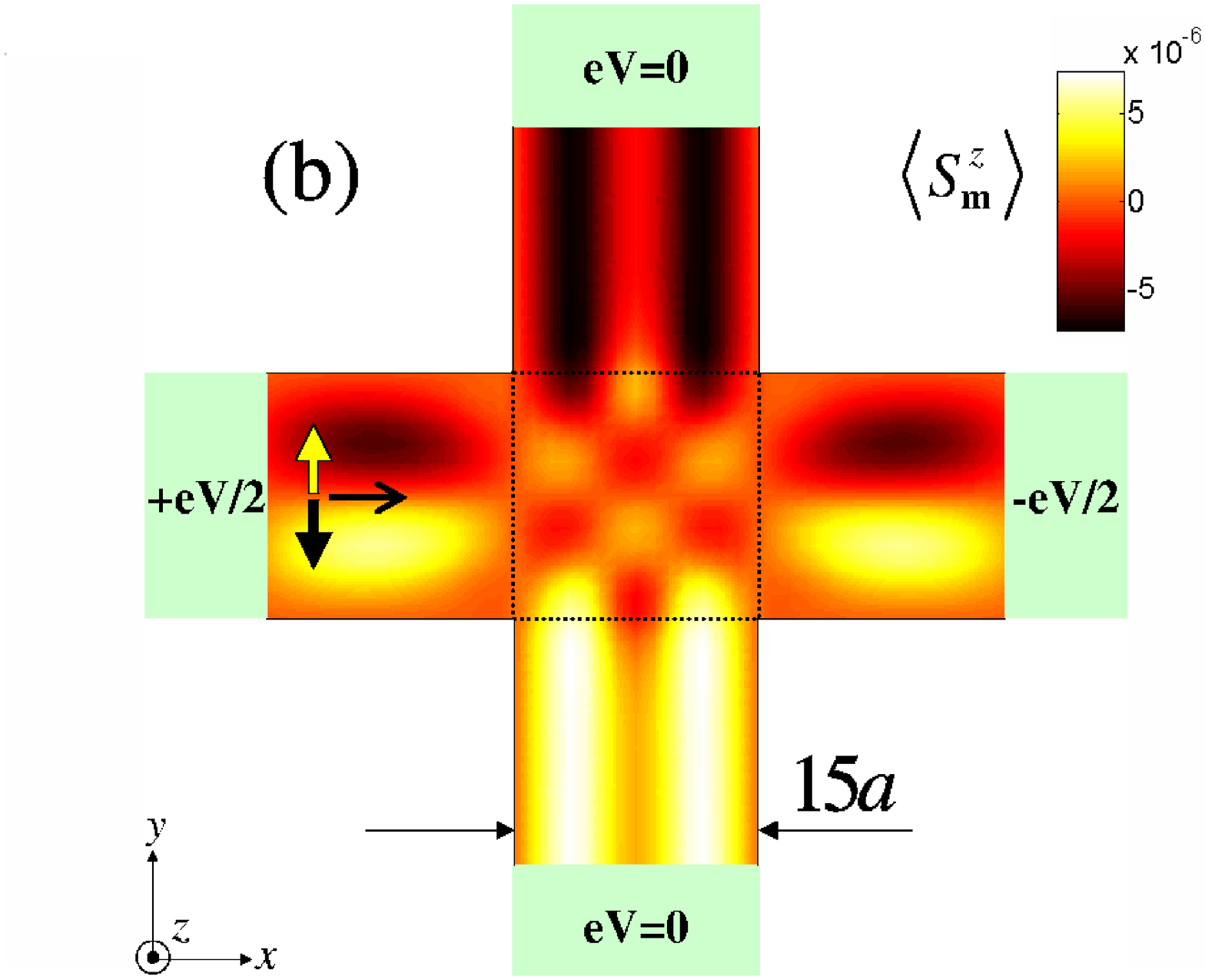,scale=0.38,angle=-90} \hspace{0.5in} \psfig{file=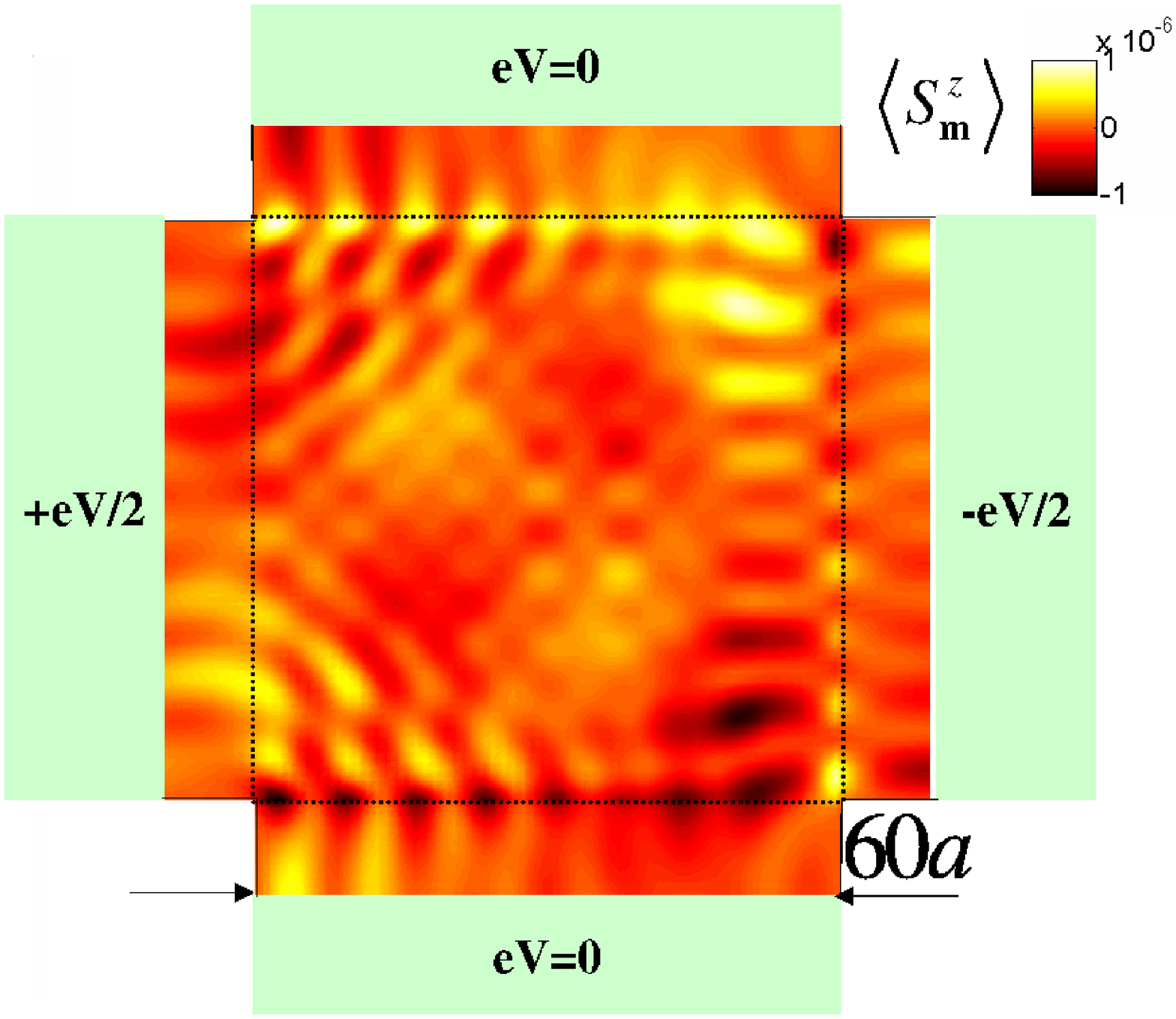,scale=0.38,angle=-90}}
\caption{(Color online) (a) The spatial distribution of the nonequilibrium local spin current $\left< \hat{J}_{\bf mm'}^{S_z({\rm neq})} \right>$ and (b) the flowing spin density $\left<\hat{S}_{\bf m}^z \right>$ 
in ballistic bridges consisting of four ideal leads ($t_{\rm SO} \equiv 0$) attached to a finite-size 2DEG with the Rashba SO  coupling $t_{\rm SO}=0.1t_{\rm o}$ and the corresponding spin precession length $L_{\rm SO} \approx 15.7a$. The 2DEG central region is of the size $15a \times 15a \approx L_{\rm SO} \times L_{\rm SO}$ in the left column; and $60a \times 60a$ in the right column. Note that the linear response ($eV=10^{-3}t_{\rm o}$) total transverse spin Hall current $I^s_{\rm trans}(m_y) = \sum_{m_x} \left <\hat{J}_{(m_x,m_y)(m_x,m_y+1)}^{S_z}\right>$ in the leads ($m_y \le 10$ or $m_y \ge 61$) flows along the negative $y$-axis from (i.e., from the top to the bottom transverse lead) in both cases.}\label{fig:macroscopic}
 \end{figure*}

To highlight mesoscopic features~\cite{baranger1989a} (such as the effect of the measuring geometry and the properties of the attached probes) of the spin Hall effect in multiterminal  ballistic SO coupled structure,~\cite{nikolic_mesoshe} we plot in Fig.~\ref{fig:adiabatic}  the spatial profile of microscopic   spin currents for the four-terminal bridge attached to the leads where Rashba SO coupling  is switched  on adiabatically (via a linear function) within the finite region of the leads adjacent to  the 2DEG  sample in the center of the device. In this measuring setup, the reflection~\cite{nikolic_purity} at the  interface separating zero and non-zero SO coupling regions 
is greatly suppressed, thereby enhancing the spin Hall  current~\cite{nikolic_mesoshe} [as encoded by 
longer arrows in the profiles of Fig.~\ref{fig:adiabatic} when compared to Fig.~\ref{fig:eq_vs_neq}(d)]. 

\begin{figure*}
\centerline{\psfig{file=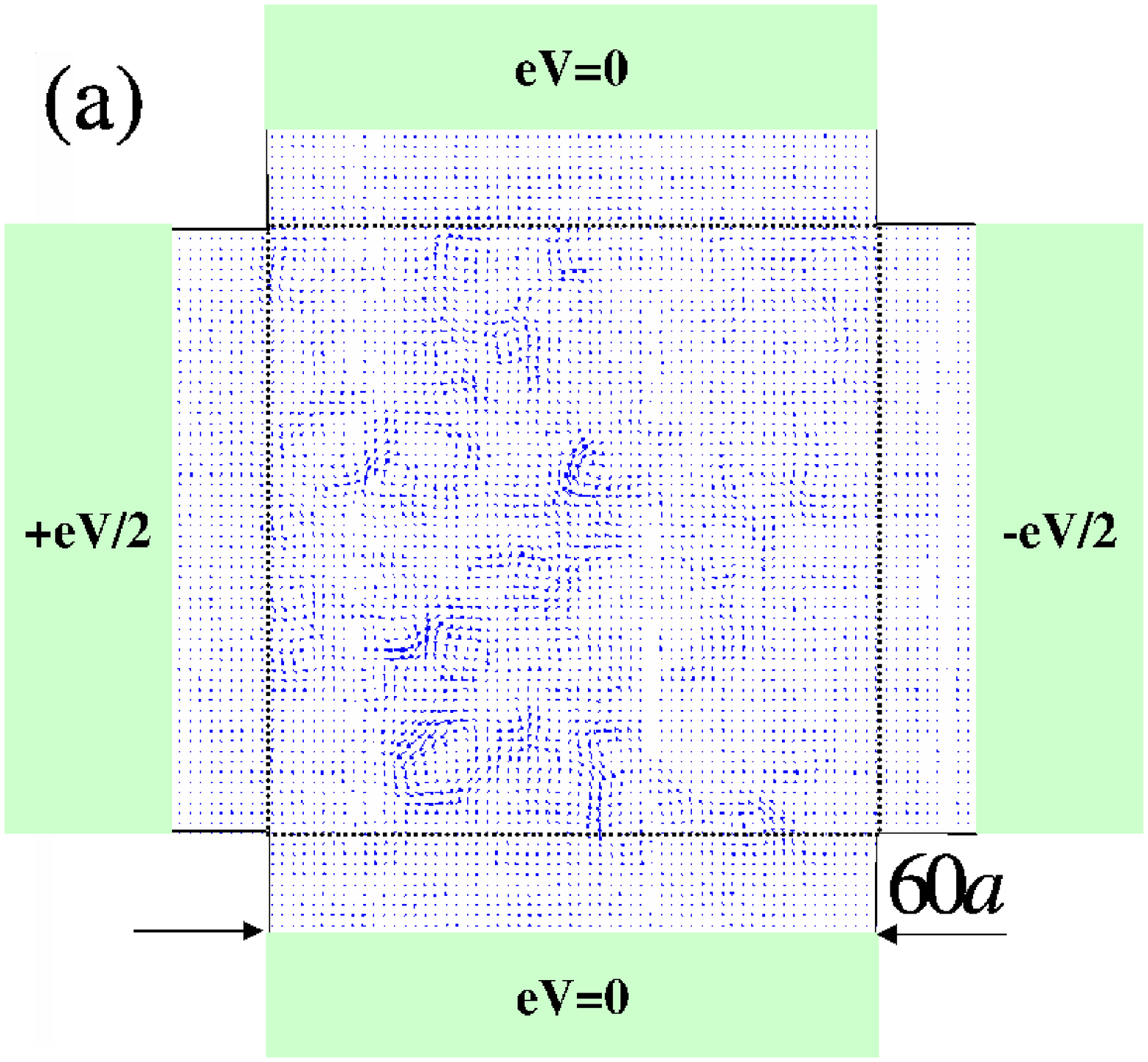,scale=0.38,angle=-90} \hspace{0.3in} \psfig{file=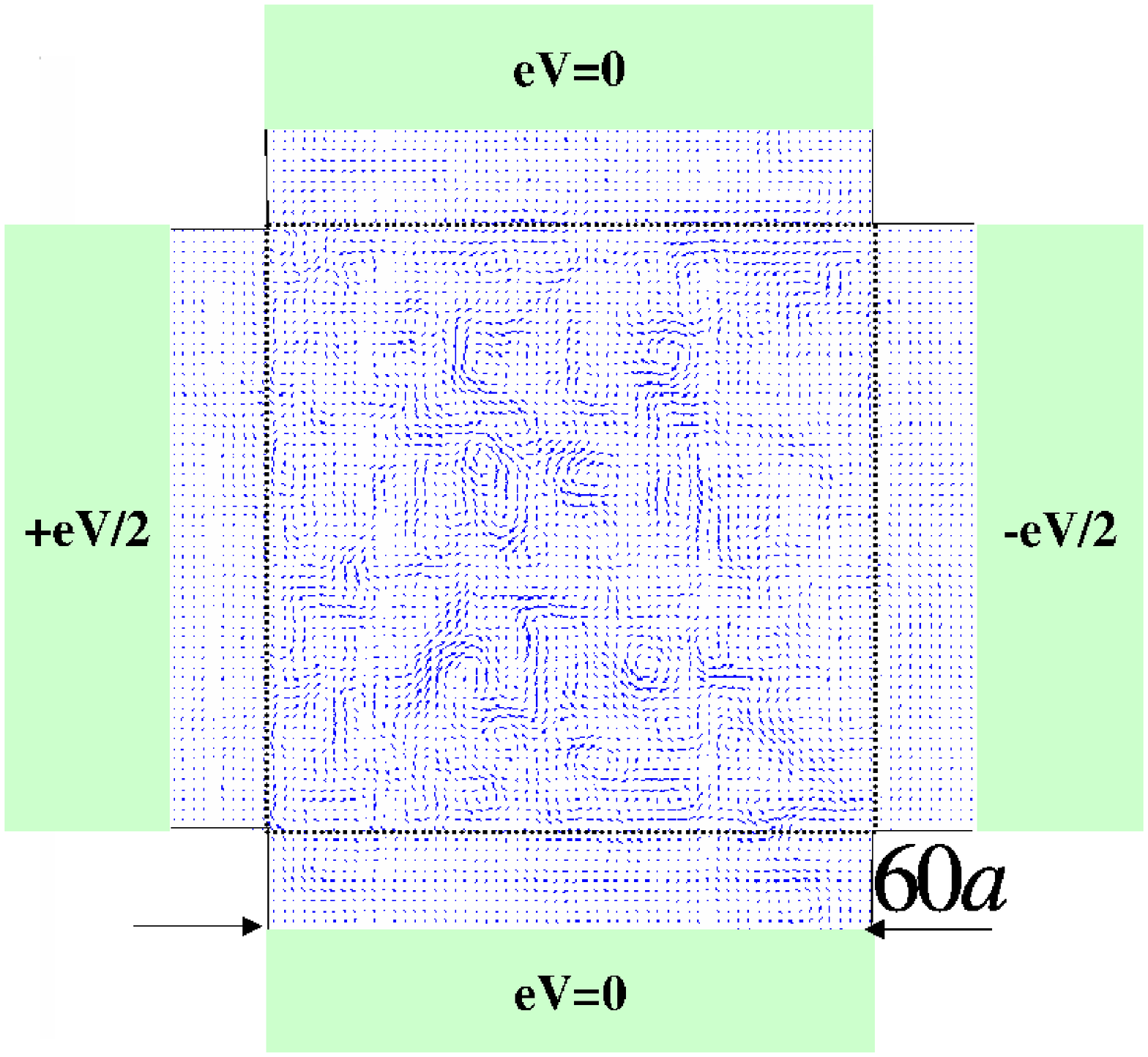,scale=0.38,angle=-90}} \vspace{0.5in}
 \centerline{\psfig{file=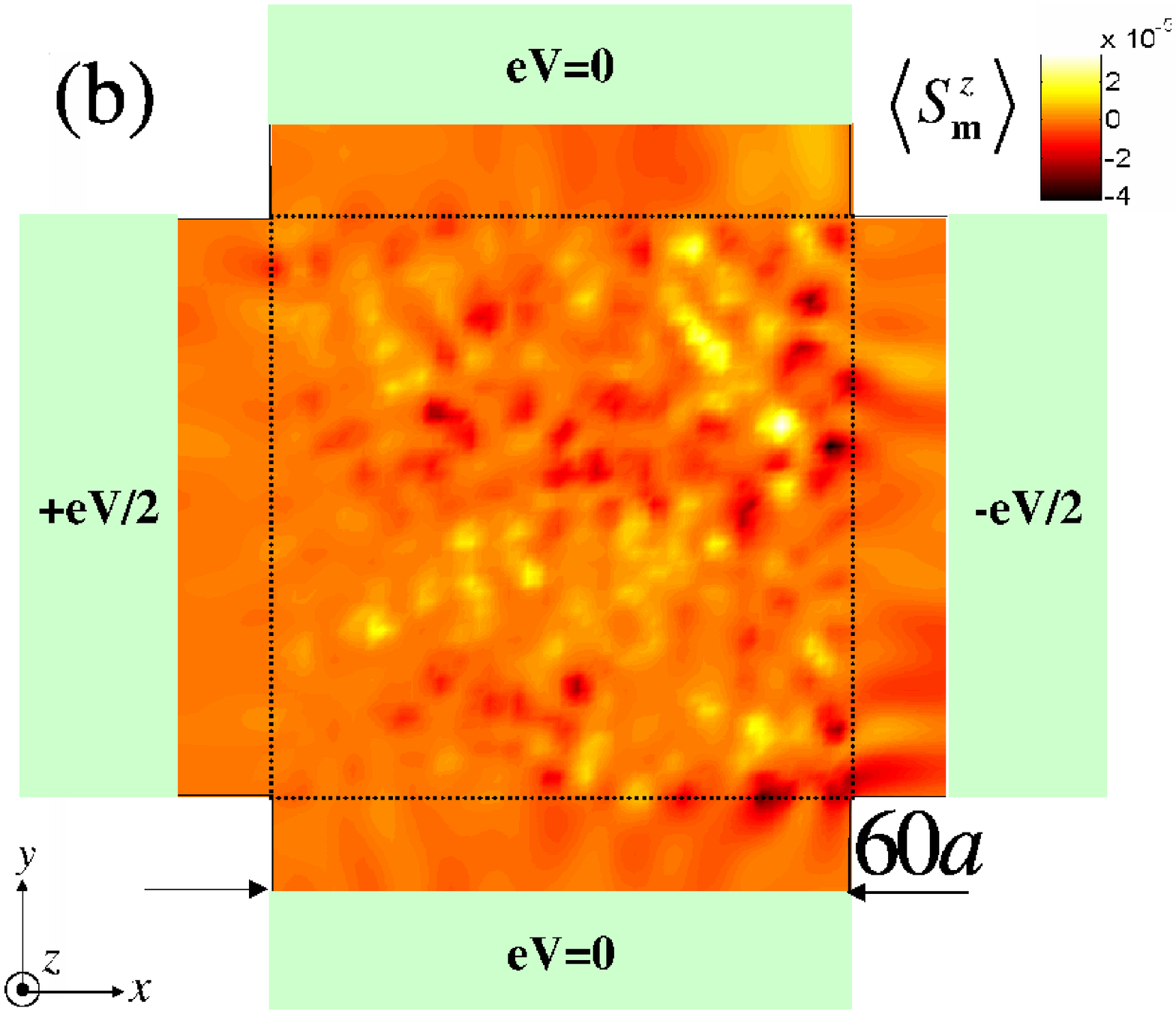,scale=0.38,angle=-90} \hspace{0.3in} \psfig{file=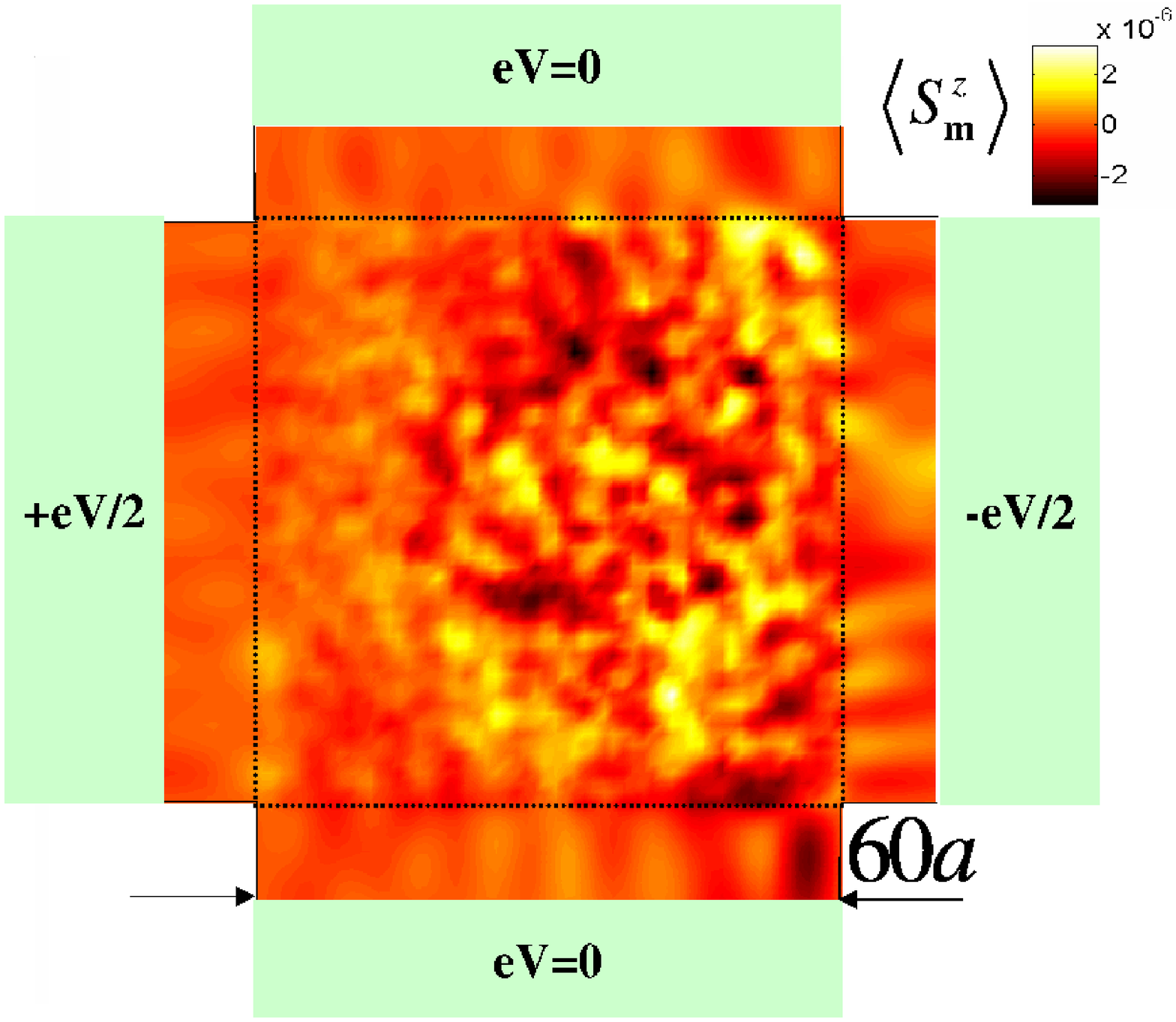,scale=0.38,angle=-90}}
\caption{(Color online) (a) The spatial distribution of the nonequilibrium local spin current $\left< \hat{J}_{\bf mm'}^{S_z({\rm neq})} \right>$ and (b) the flowing spin density $\left<\hat{S}_{\bf m}^z \right>$ 
in the disordered 2DEG of the size $L=60 a$ with static spin-independent impurities which set the mean 
free path $\ell =7a$. The Rashba SO  coupling within the 2DEG region is $t_{\rm SO}=0.1t_{\rm o}$ ($t_{\rm SO} \equiv 0$ in ideal the leads) and the corresponding D'yakonov-Perel' spin relaxation length~\cite{zutic2004a} is $L_{\rm SO} \approx 15.7a$. 
The applied bias voltage $eV=10^{-3}t_{\rm o}$ between the longitudinal leads drives the linear 
response longitudinal charge transport in the diffusive regime ($L \gg \ell$). Panels in the left column correspond to a single sample with specific configuration of impurities, while panels in the right column show disorder-averaged profiles over an ensemble of  100 different impurity configurations.}\label{fig:diffusive}
 \end{figure*}

Since local spin current  within the central  region is not conserved, as manifested by the total spin 
current changing magnitude between different transverse cross sections (see  Sec.~\ref{sec:vs} for total spin current profiles) separated by long $\gtrsim L_{\rm SO}$, the 
spatial profiles of the local spin current density appear not to be directly measurable.~\cite{rashba_review} However, the nonequilibrium spin density (i.e., the related spin magnetization) is a well-defined and measurable quantity.~\cite{kato2004a,wunderlich2005a} Therefore, we plot the spatial distribution of the stationary flowing  $\left<\hat{S}_{\bf m}^z \right>$ Eq.~(\ref{eq:spin_density}) spin density in Fig.~\ref{fig:adiabatic} to gain an additional insight into the dynamics of the spin Hall quantum transport. These microscopic picture convincingly demonstrate how spin-$\uparrow$ and spin-$\downarrow$ densities flow in opposite transverse directions through the attached leads, without any net charge flow in the transverse direction, thereby giving rise to a {\em pure} 
transverse  spin Hall current. When we reverse the direction of the longitudinal charge current (by reversing 
the bias voltage $V \rightarrow -V$), the transverse spin current and spin densities flip their sign, 
as exploited in experiments to confirm strong signatures of the spin Hall effect.~\cite{wunderlich2005a} 

Thus, in contrast to the arguments~\cite{zhang2005a} advocating impossibility of spin transport and accumulation by mechanisms driven solely by the intrinsic SO coupling terms in the effective Hamiltonian of 
spin-split semiconductors, Fig.~\ref{fig:adiabatic} demonstrates that spin Hall effect originating in 
ballistic multi-terminal  devices  (without necessity for impurity induced effects) generates 
genuine nonequilibrium spin flux that can be used for spin injection and spintronics applications.~\cite{zutic2004a}

Another mesoscale driven property of the spin Hall effect in ballistic nanostructures is its finite-size scaling being governed by the processes on the spin precession length $L_{\rm SO}$: One can 
differentiate the ``mesoscopic'' regime $L \lesssim L_{\rm SO}$ where spin Hall current oscillates with 
increasing 2DEG size (changing sign with increasing size $L \times L$ of the 2DEG for strong enough 
Rashba SO coupling $t_{\rm SO} \gtrsim  0.04t_{\rm o}$); and the ``macroscopic'' regime $L \gg L_{\rm SO}$ where it saturates at some average value.~\cite{nikolic_mesoshe} While the spatial profiles of local spin currents and spin densities are easy to interpret for $L < L_{\rm SO}$ (as in Fig.~\ref{fig:eq_vs_neq}),  Fig.~\ref{fig:macroscopic} divulges how they become increasingly more intricate in the samples of the size $L_{\rm SO} \times L_{\rm SO}$ (for which the spin Hall current reaches its maximum~\cite{nikolic_mesoshe}), or in the macroscopic regime $L \gg L_{\rm SO}$. Nevertheless, the total spin Hall current exists only in 
the transverse leads (the sum of bond $S_z$-spin currents  over the cross section of longitudinal leads is equal to zero) with its magnitude and sign being identical to the terminal spin currents obtained within the Landauer-B\" uttiker formalism~\cite{nikolic_mesoshe} (see Sec.~\ref{sec:vs} for quantitative comparison).

\section{Bulk vs. edge local spin currents in Disordered Four-Terminal 
Rashba coupled Nanostructures}\label{sec:disorder}

A surprising feature of the early predictions for the intrinsic spin Hall effect in hole-doped~\cite{murakami2003a} or electron-doped~\cite{sinova2004a} infinite homogeneous 
SO coupled semiconductor systems is apparent insensitivity of the spin Hall conductivity 
$\sigma_{sH}$ on the mean free  path and relaxation rates.~\cite{zhang2005a} This has 
provoked intense theoretical scrutiny of the effects which spin-independent scattering off 
static impurities~\cite{schliemann2004a} imposes on the intrinsic spin Hall current, leading to 
the conclusion that, in fact, $\sigma_{sH} \rightarrow 0$ vanishes~\cite{inoue2004a,sheng2005b} 
for arbitrary small disorder in any model with SO coupling linear in momentum (such as the 
Rashba Hamiltonian of the spin-slit 2DEG) due to accidental cancellations. In the general case,  
where SO coupling contains higher order momentum terms,~\cite{murakami2004a,malshukov2005a} 
$\sigma_{sH}$ can be resilient to sizable disorder strengths.~\cite{chen2005a} Moreover, 
an influential conjecture has emerged from the studies of the disorder effects on the 
intrinsic spin Hall current---macroscopic inhomogeneities facilitate spin currents~\cite{rashba_review,shekhter2005a} so  that transverse {\em edge} spin  
current ${\mathcal J}_y^z$ could emerge near the sample-electrode interface,~\cite{shytov2004a} 
even in systems where it is expected to be destroyed in the bulk.~\cite{inoue2004a}

However, the quantitative support for the picture of edge spin Hall currents is based on the 
analysis of semiclassical diffusive transport through a rather abstract structure, where 
2DEG infinite in the transverse  direction is attached to two massive electrodes in the 
longitudinal direction,~\cite{shytov2004a} that does not correspond to any experimentally 
realizable device. On the other hand, the presence of SO couplings makes the dynamics of 
transported spin in experimentally  relevant confined structures strongly dependent on the 
properties of their interfaces, boundaries, and the attached electrodes,~\cite{nikolic_purity} 
even for semiclassical spatial propagation of charges which carry spins evolving according to 
quantum dynamical laws.~\cite{chang2004a} For example, heuristic arguments based on the 
Keldysh formalism applied to an infinite two-terminal structure (lacking actual lateral edges) 
of Ref.~\onlinecite{shytov2004a} suggest that nonequilibrium spin Hall accumulation $\left< S^z_{\bf m} \right> \neq 0$ will appear only in the four corners at the lead/2DEG interfaces (due  to ${\mathcal J}_y^z \neq 0$ existing within a spin relaxation length $L_{\rm SO}$ wide region around such interfaces), 
in contrast to the Keldysh formalism applied to finite-size 2DEG in the Landauer two-terminal setup 
where non-zero spin accumulation (with opposite sign on the two lateral edges~\cite{kato2004a,wunderlich2005a}) is found along the whole lateral edge.~\cite{nikolic_accumulation,onoda2005a}

\begin{figure}
\centerline{\psfig{file=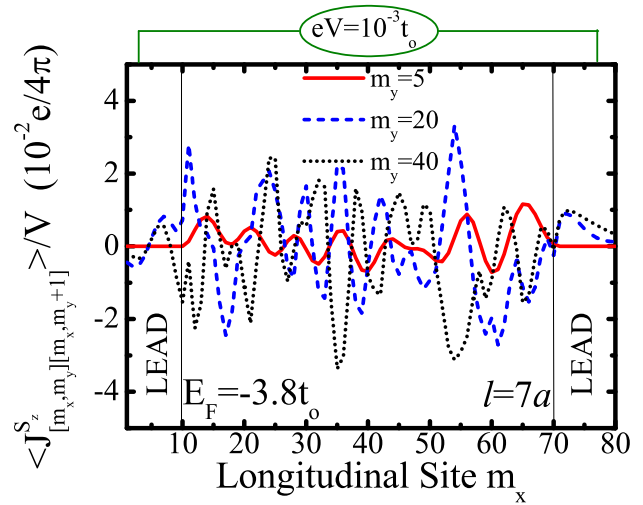,scale=0.8,angle=0}} 
\caption{(Color online) The  longitudinal profile of the disorder-averaged 
(over 100 samples) nonequilibrium bond spin currents, whose full two-dimensional 
profile is shown in the right panel of Fig.~\ref{fig:diffusive}(a), across the transverse 
cross  sections in the ideal bottom lead ($m_y=5$) and inside the diffusive 2DEG 
($m_y=20$ and $m_y=40$) with static spin-independent impurities. Note that the length of the $m_y=5$ 
cross section within the bottom transverse lead is $60a$, while the length of $m_y=20$ 
and $m_y=40$ transverse cross sections is $80a$ ($60a$ inside the 2DEG + $10a$ within 
each of the two longitudinal leads).}\label{fig:disorder_profile}
\end{figure}

Within the formalism of bond spin currents of Sec.~\ref{sec:bond_spin} (which represent the lattice version of the same quantity ${\mathcal J}^z_y$ studied in Ref.~\onlinecite{shytov2004a}) these issues can be resolved through the exact evaluation of the retarded and lesser Green functions a non-interacting particle propagating through 
the random potential in finite-size multiterminal SO  coupled device. Thus, we plot in Fig.~\ref{fig:diffusive} the spatial distribution of  the local spin currents and  spin  densities 
for a single disordered Rashba SO coupled 2DEG, as well as their disorder-averages (which are to be compared 
with the analysis based on the diffusion equation~\cite{shytov2004a}). The disorder strength is tuned 
to ensure the diffusive transport regime $\ell \ll L$, while the magnitude of the spin Hall current 
in the leads at this concentration of impurities is still about $~80\%$ of its maximum value set in 
the clean limit for the same four-terminal nanostructure.~\cite{nikolic_mesoshe} We do not find any evidence for the confinement of spin fluxes near the 2DEG/longitudinal-leads interfaces. The conclusion based on the spatial profile of microscopic spin currents [i.e., the  ``distribution of arrows'' in  Fig.~\ref{fig:diffusive}(a)] is further corroborated in Fig.~\ref{fig:disorder_profile} by the one-dimensional longitudinal profiles of the bond spin currents on different cross sections within  the SO coupled sample and in the leads. Thus, both Fig.~\ref{fig:diffusive} and Fig.~\ref{fig:disorder_profile} suggest that precessing spins will propagate through the bulk of the diffusive Rashba SO coupled 2DEG in semiconductor heterostructure.

\begin{figure}
\centerline{\psfig{file=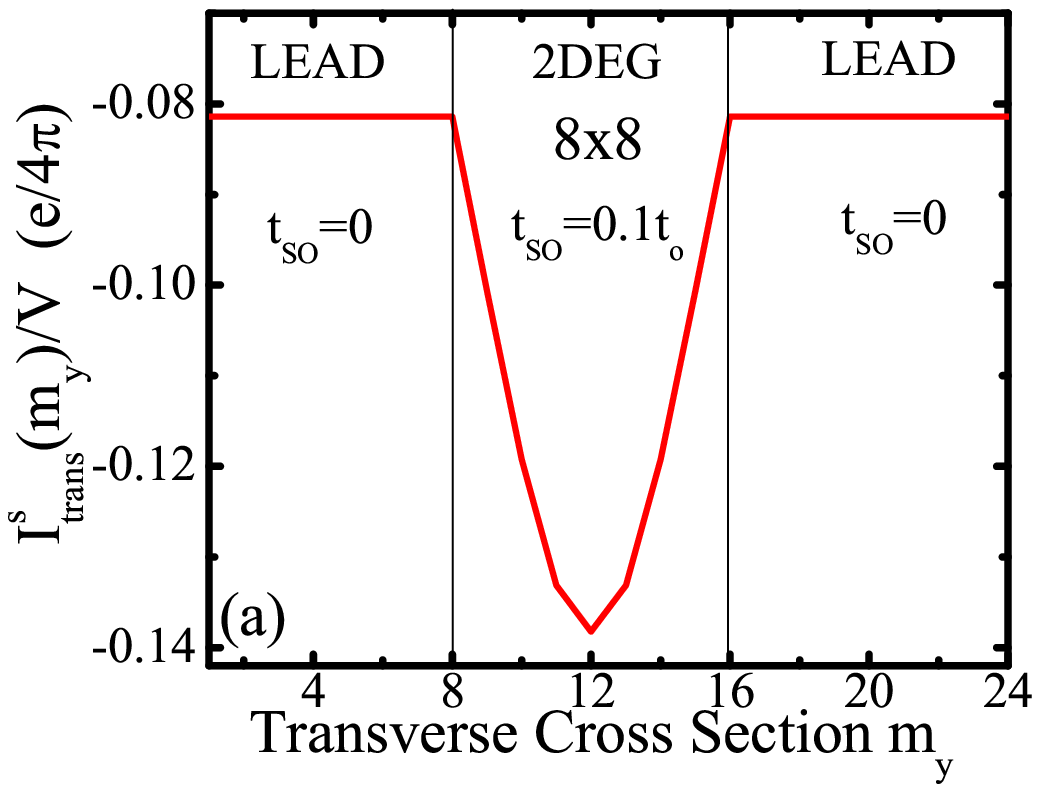,scale=0.59,angle=0}} 
\centerline{\psfig{file=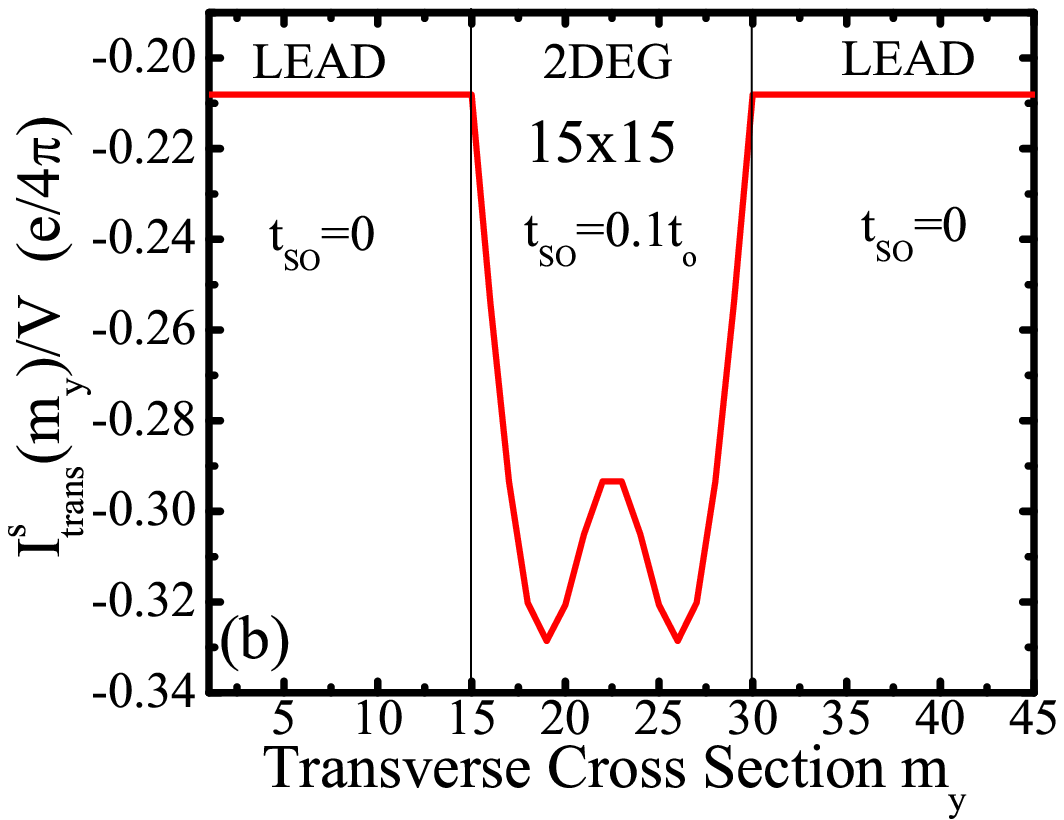,scale=0.59,angle=0}}
\centerline{\psfig{file=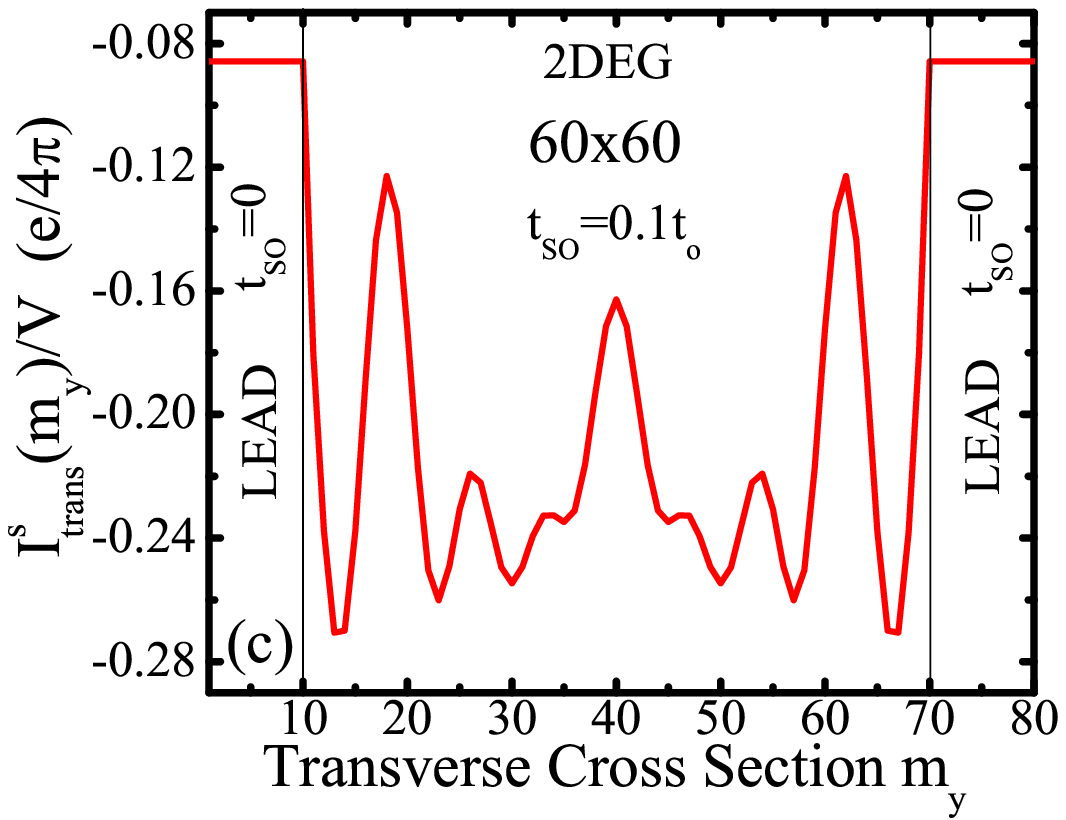,scale=0.59,angle=0}}
\caption{(Color online) The total spin current $I^s_{\rm trans}(m_y)$ on 
transverse cross sections across the ideal bottom and top transverse 
leads (with no SO coupling) and the Rashba SO coupled 2DEG with  
$t_{\rm SO}=0.1t_{\rm o}$ ($L_{\rm SO} \approx 15.7a$). The linear  response 
longitudinal charge transport (at zero temperature) is driven by the applied 
bias  voltage $eV=10^{-3} t_{\rm o}$ between the longitudinal leads of the 
four-terminal structures where unpolarized electrons are injected from the 
left longitudinal lead at the Fermi energy $E_F=-3.8t_{\rm o}$. The total pure 
transverse spin current in panel (a) is obtained by summing the bond spin 
currents in Fig.~\ref{fig:eq_vs_neq}(d) over a cross section $m_y$; similarly, panel (b) and 
(c) corresponds to such sums for the profiles in the left and right columns of 
Fig.~\ref{fig:macroscopic}, respectively.}\label{fig:total_trans}
\end{figure}

\begin{figure}
\centerline{\psfig{file=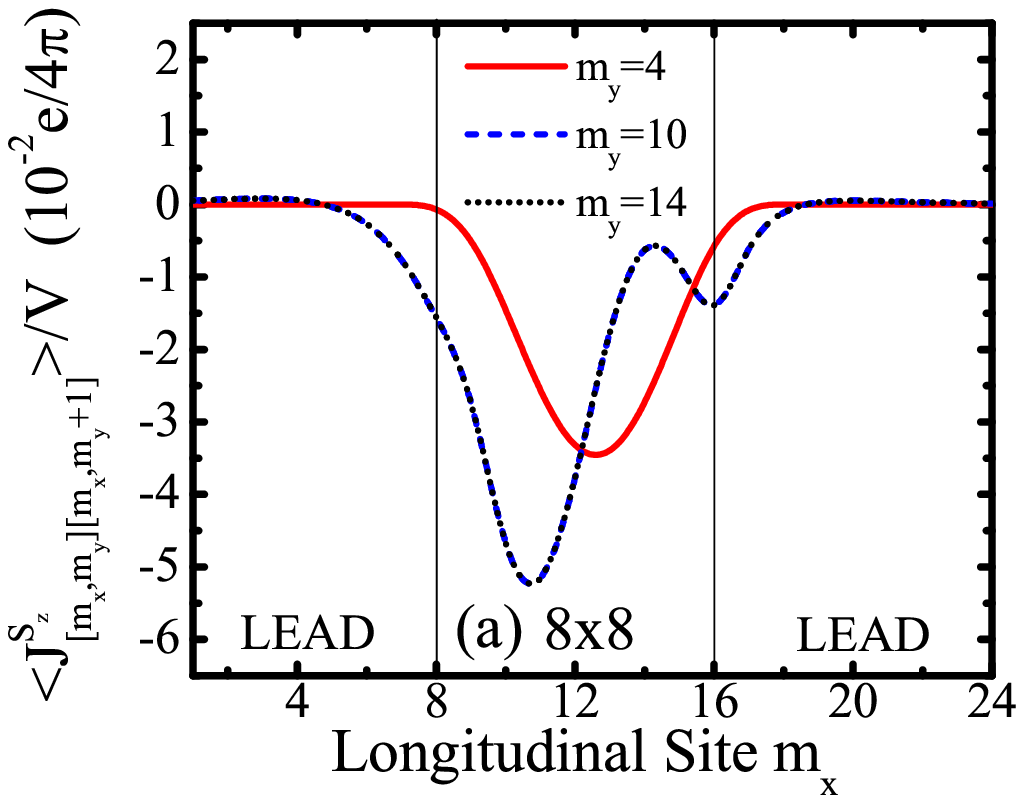,scale=0.6,angle=0}} 
\centerline{\psfig{file=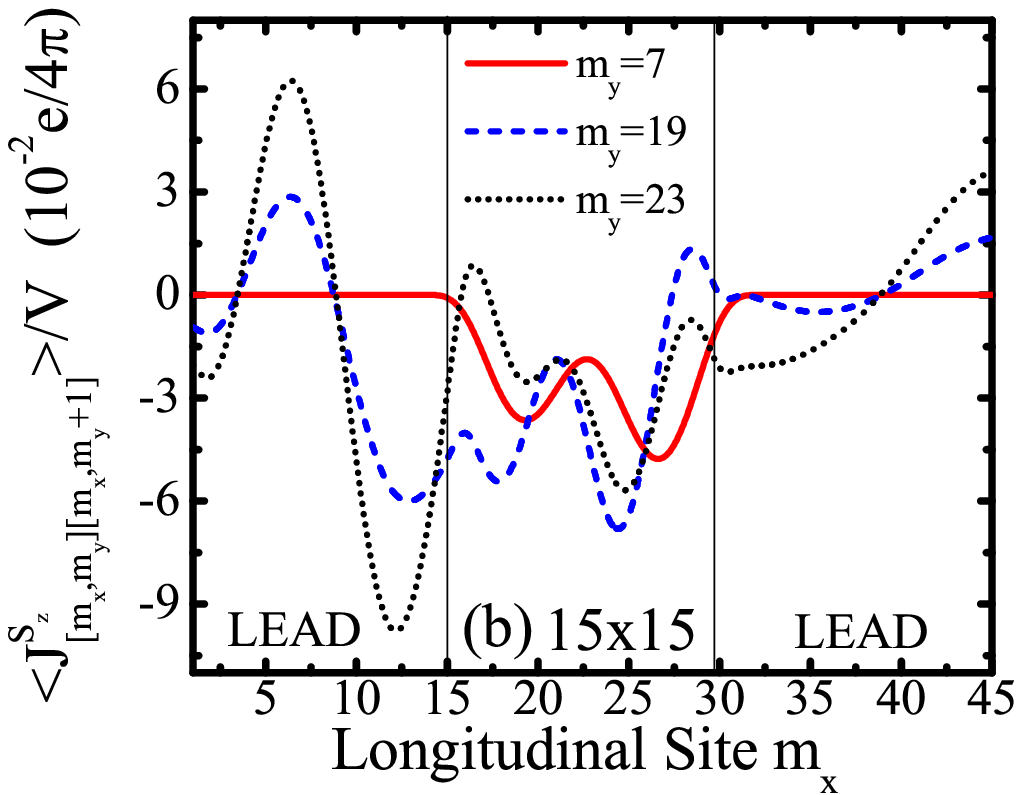,scale=0.6,angle=0}}
\centerline{\psfig{file=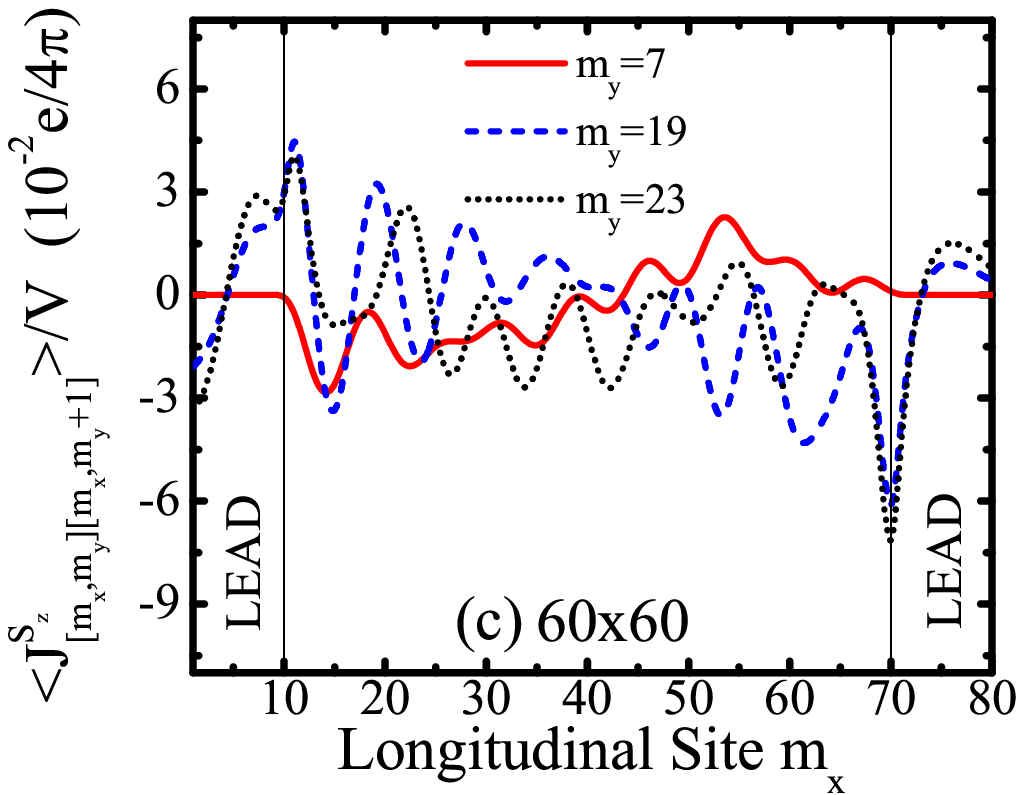,scale=0.6,angle=0}}
\caption{(Color online) The  longitudinal profile of the  nonequilibrium bond spin currents  along 
the transverse cross sections $m_y$ in the ideal bottom lead (solid curves) and inside 
the SO coupled ballistic 2DEG (dashed and dotted curves).  The Rashba SO coupling is $t_{\rm SO} =0.1 t_{\rm o}$ ($L_{\rm SO} \approx 15.7a$) within the central 2DEG region and $t_{\rm SO} \equiv 0$ in both the longitudinal and the transverse leads of the four-terminal bridge. The sum of the longitudinal profiles gives the total transverse spin Hall current in Fig.~\ref{fig:total_trans}. Note that profile in panel 
(a) corresponds to bond spin current magnitude represented by arrows in 
Fig.~\ref{fig:eq_vs_neq}(d) and, similarly, panels (b) and (c) corresponds to the left and 
right columns in Fig.~\ref{fig:macroscopic}, respectively.}\label{fig:long_profile}
\end{figure}

\section{Total spin Hall current vs. local spin Hall current: Landauer-B\" uttiker vs. Landauer-Keldysh picture} \label{sec:vs}

The traditional charge transport experiments measure total current $I$ and the conductance $I=GV$ 
relating  the charge current to the voltage drop $V$, rather than local current density ${\bf j}$ and 
the  conductivity ${\bf j} =\sigma {\bf E}$ relating it to the externally applied electric field 
(note also that in ballistic transport or quantum-coherent transport  through a diffusive conductor  
conductivity ceases to exist as a local quantity~\cite{baranger1989a}). Since realization of 
the total pure spin currents has been detected experimentally in optical pump-probe experiments,~\cite{stevens2003}  and several theoretical schemes are proposed to detect 
them indirectly via various electrical measurements,~\cite{hirsch1999a,hankiewicz2004a,meier2003a} 
we focus in this section on the properties of total spin Hall current $I^s_{\rm trans}(m_y)$ on 
different transverse cross sections of four-terminal devices, which is obtained from Eq.~(\ref{eq:total_trans}) by summing the bond spin currents.

As shown in Fig.~\ref{fig:total_trans}, the total pure spin current in the transverse 
leads, obtained by summing the nonequilibrium bond spin currents $\left<\hat{J}_{\bf mm'}^{S_z(\rm neq)}\right>$ over an  arbitrary transverse cross section of the ideal leads (where SO coupling
vanishes), flows through them in a conserved fashion, $I_{\rm trans}^s(m_y) = {\rm const.}$ for 
any $m_y \in {\rm lead}$. However, the same summation over  the transverse cross sections within  
the 2DEG yields a quantity which is not conserved, except on the short length scales $\ll L_{\rm SO}$. 
This is due to the fact that, e.g., injected eigenstate $|\!\! \uparrow \rangle$ of $\hat{\sigma}_z$ 
will precess in the effective magnetic field of the Rashba SO coupling (along the $y$-axis), thereby  
changing the amplitude of the spin current measured with respect to the $z$-axis as the spin 
quantization axis. Additional quantitative information about the microscopic details of spin fluxes 
is provided by Fig.~\ref{fig:long_profile} which plots the one-dimensional longitudinal profiles of 
the bond spin currents over the selected transverse cross sections (in the leads and in the 2DEG 
sample) cutting through the full spatial distributions of Fig.~\ref{fig:eq_vs_neq} and 
Fig.~\ref{fig:macroscopic}. Note that the sum of these longitudinal profiles yields the corresponding 
total spin current at the cross section $m_y$ in Fig.~\ref{fig:total_trans}.

\begin{figure}
\centerline{\psfig{file=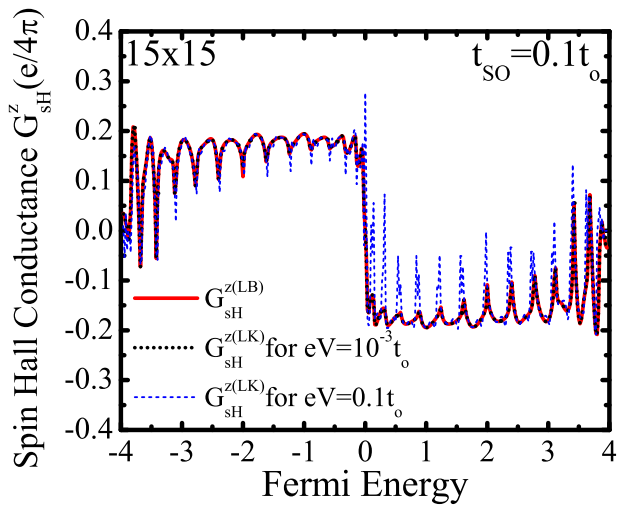,scale=0.8,angle=0}} 
\caption{(Color online) The spin Hall conductance of a Rashba SO coupled four-terminal 2DEG, obtained from the Landauer-B\" uttiker  linear response transmission formalism as $G_{sH}^{z({\rm LB})} = \lim_{V \rightarrow 0} I^s_2 /V$ [where $I_2^s$  denotes the terminal spin current in the top transverse lead in Fig.~\ref{fig:macroscopic}(a)], vs. the spin Hall conductance  $G_{sH}^{z(\rm LK)} = I^s_{\rm trans}(m_y)/V$ obtained within the Landauer-Keldysh formalism by summing the bond spin currents  over the transverse cross section in the top lead of structure in Fig.~\ref{fig:macroscopic}(a) for low bias voltage $eV =10^{-3}t_{\rm o}$ or high bias voltage $eV =0.1 t_{\rm o}$ applied between the longitudinal leads. The size of the central 2DEG region is $15a \times 15 a \approx L_{\rm SO} \times L_{\rm SO}$, for which the spin Hall conductance $G_{sH}^{z({\rm LB})}(L,t_{\rm SO})$ reaches maximum when increasing the sample size at fixed SO coupling.~\cite{nikolic_mesoshe}}\label{fig:lb}
\end{figure}

The total spin currents in the leads in the linear response regime $V \rightarrow 0$ can also be calculated using the spin-dependent Landauer-B\" uttiker scattering formalism for spin currents in multiprobe geometries.~\cite{nikolic_mesoshe,souma_ringshe,pareek2004a} In this formalism, one has to compute the 
spin-resolved transmission matrices connecting the spin-polarized asymptotic scattering states in 
semi-infinite ideal  leads attached to the sample,~\cite{nikolic_purity} which requires the knowledge 
of only the elements of the  retarded Green function matrix Eq.~(\ref{eq:retarded}) between the sites of 
the sample which are adjacent to the leads (similar expressions in terms of only the retarded Green 
function between the sample edges are obtained for the total charge 
currents~\cite{caroli1971a,nonoyama1998a}). We demonstrate in Fig.~\ref{fig:lb} that the spin Hall 
conductance obtained from the Landauer-B\" uttiker formalism $G_{sH}^{z({\rm LB})} = \lim_{V \rightarrow 0} I^s_2 /V$ (where 2 denotes the top transverse lead~\cite{nikolic_mesoshe}) is almost identical to the spin Hall 
conductance $G_{sH}^{z({\rm LK})} = I^s_{\rm trans}(m_y)/V$ obtained by summing the bond spin currents over a 
cross section $m_y$ in the top transverse ideal lead, on the proviso that the applied bias voltage is 
small $eV \ll E_F$ and current is carried only by the states at the Fermi level. This result further 
justifies the introduction of nonequilibrium bond spin current formula Eq.~(\ref{eq:bond_spincurrent_kin}) 
for the $z$-component of spin.

\section{Concluding Remarks}\label{sec:conclusion}

In conclusion, we have demonstrated how to define the bond spin current, describing the spin flux 
across a single bond between two sites of the lattice model of an SO coupled semiconductors, and 
evaluate it in terms of the Keldysh nonequilibrium Green functions for to the Landauer setup 
where finite-size sample is attached to many semi-infinite ideal leads to form a theoretical model 
of experimentally accessible spin Hall bridges. Although spin current is not conserved within the SO coupled region (i.e., on the length scales comparable to the spin precession length $L_{\rm SO}$ on which the SO coupling manifests itself), the microscopic spin fluxes are nearly conserved on short scales. Thus, the bond spin 
currents make it possible to obtain their spatial distribution by following the dynamics of transported 
spin on the scale of the lattice spacing $a \ll L_{\rm SO}$. Such profiles of (the lattice 
version of) local spin current density, together with stationary flow profiles of  physically transparent 
(and measurable) local spin densities, allows us to demonstrate microscopic details of how pure transverse 
spin Hall current emerges in {\em clean} Rashba SO coupled 2DEG through which the unpolarized longitudinal 
charge current flows ballistically (where electrons do not scatter off impurities and do not feel the 
electric field). These spatial profiles are highly dependent on whether the 2DEG size is smaller or greater 
than the spin precession length, and can be affected by non-trivial measurement geometries, as discussed 
in the theory of the mesoscopic spin Hall effect.~\cite{nikolic_mesoshe} The bond spin current within 
the bulk of the 2DEG is also resilient to weak disorder so that spin fluxes remain non-zero in the bulk of 
the sample and are not localized near the edges of the diffusive Rashba spin-split 2DEG. 

Using the bond spin current we explicitly demonstrate that nonequilibrium total spin current can be  carried 
only by the states around the Fermi energy, while the Fermi sea contributes to local persistent spin currents  which, however, do not transport any spin through a given cross section. The nanometer scale details of the spin Hall flow in multiterminal mesoscopic (quantum-coherent) structures convincingly show that the intrinsic to the crystal SO couplings can be used to generate spin fluxes, spin accumulation, and ultimately be employed to construct all-electrical spin injectors which do no require any ferromagnetic elements (whose coupling to semiconductors has been one of the major impediments for spintronics applications~\cite{zutic2004a}).

\begin{acknowledgments}
We are grateful to J. Inoue and J. Sinova for valuable discussions. Acknowledgment is made to the donors 
of the American Chemical Society Petroleum Research Fund for partial support of this research.
\end{acknowledgments}

\end{document}